\documentclass{ieeeaccess}
\usepackage[nocompress]{cite}
\usepackage{amsmath,amssymb,amsfonts}
\usepackage{algorithmic}
\usepackage{graphicx}
\usepackage{textcomp}
\usepackage{booktabs}
\usepackage{multirow}
\usepackage{url}

\usepackage{bm}
\definecolor{urlblack}{rgb}{0,0,0}
\usepackage[colorlinks=true,linkcolor=blue,citecolor=blue,urlcolor=urlblack,bookmarks=false]{hyperref}

\makeatletter
\DeclareRobustCommand{\eqref}[1]{\hyperref[#1]{\textup{(\ref*{#1})}}}
\AtBeginDocument{\DeclareMathVersion{bold}
\SetSymbolFont{operators}{bold}{T1}{times}{b}{n}
\SetSymbolFont{NewLetters}{bold}{T1}{times}{b}{it}
\SetMathAlphabet{\mathrm}{bold}{T1}{times}{b}{n}
\SetMathAlphabet{\mathit}{bold}{T1}{times}{b}{it}
\SetMathAlphabet{\mathbf}{bold}{T1}{times}{b}{n}
\SetMathAlphabet{\mathtt}{bold}{OT1}{pcr}{b}{n}
\SetSymbolFont{symbols}{bold}{OMS}{cmsy}{b}{n}
\renewcommand\boldmath{\@nomath\boldmath\mathversion{bold}}}
\makeatother

\def\BibTeX{{\rm B\kern-.05em{\sc i\kern-.025em b}\kern-.08em
    T\kern-.1667em\lower.7ex\hbox{E}\kern-.125emX}}

\usepackage{arxiv-layout}

\begin{document}

\title{Connectionless Bluetooth Channel Sounding via PAwR for Scalable and Energy-Efficient Ranging}
\author{\uppercase{Leon Schex}\authorrefmark{1}, \IEEEmembership{Graduate Student Member, IEEE},
\uppercase{Markus Cremer}\authorrefmark{1}, AND
\uppercase{Uwe Dettmar}\authorrefmark{1}, \IEEEmembership{Member, IEEE}}

\address[1]{Institute of Computer and Communication Technology (ICCT), TH Köln -- University of Applied Sciences, 50679 Cologne, Germany (e-mail: leon.schex@th-koeln.de; markus.cremer@th-koeln.de; uwe.dettmar@th-koeln.de)}

\markboth
{Schex \headeretal: Connectionless Bluetooth Channel Sounding via PAwR for Scalable and Energy-Efficient Ranging}
{Schex \headeretal: Connectionless Bluetooth Channel Sounding via PAwR for Scalable and Energy-Efficient Ranging}

\corresp{Corresponding author: Leon Schex (e-mail: leon.schex@th-koeln.de).}
\tfootnote{This work has been submitted to the IEEE for possible publication. Copyright may be transferred without notice, after which this version may no longer be accessible.}

\begin{abstract}
Bluetooth Core Specification v6.0 introduces Channel Sounding (CS) as a high-accuracy ranging primitive for Bluetooth Low Energy, but standard operation requires per-pair connections. This entails multistage initiation and limits concurrent partners per radio. We present a connectionless CS architecture combining the LE CS Test command with Periodic Advertising with Responses (PAwR). A Central Orchestrator, a gateway, and synchronized CS devices provide application-layer coordination, configuration alignment, and result aggregation. Each device derives its role, deterministic random bit generator initialization state, channel sequence, and response slot assignment from its device index and a Peer-to-Peer Assignment Matrix. The matrix can be updated per cycle to reconfigure arbitrary device-to-device pairings within each subevent group. An nRF54L15 proof-of-concept evaluation shows that deterministic channel management eliminates collision-induced outliers under simulated dense-deployment channel overlaps. Compact serialization omits fields recoverable from the measurement configuration, reducing the ranging-data payload by 68\%--69\% and enabling complete result reporting through PAwR response slots. For one measured partner per 1-s update cycle, the proposed architecture reduces steady-state active charge by up to 25\% relative to a fair connected baseline in the optional Real-time Ranging Data mode and up to 48\% in the mandatory On-demand Ranging Data mode. For a partner switch, receiving and processing a configuration payload requires approximately 97\% less charge than the connected initiation procedure. A component-based 24-h model projects up to 88\% lower charge when all four partners are replaced every update, while a timing-based capacity analysis projects up to 14\,080 active devices per PAwR train for a four-measurement workload.
\end{abstract}


\titlepgskip=-21pt

\maketitle

\section{Introduction}
\label{sec:introduction}
\PARstart{I}{ndoor} ranging and localization underpin a growing range of applications, including asset tracking, indoor navigation, secure access, and proximity services. Bluetooth Low Energy (BLE) is pervasively deployed and has long served as a low-cost baseline for proximity- and direction-based localization. Bluetooth Core Specification v6.0 introduced Channel Sounding (CS), a high-accuracy ranging primitive based on Phase-Based Ranging (PBR) and Round-Trip Time (RTT) measurements~\cite{b_core60,b_cs_overview}. PBR standardizes the multicarrier phase-difference approach to BLE phase ranging~\cite{b_nikodem2025}. Recent studies on advanced estimators, including super-resolution methods~\cite{b_santra2024} and data-driven approaches~\cite{b_tsemko2025,b_vanmarter2023,b_tariq2024}, demonstrate significant improvements in CS ranging accuracy, with data-driven methods reaching decimeter-class results even in multipath environments.

While the Bluetooth specifications define the Controller-side procedures for CS, standard CS usage in deployed systems is tied to connection-oriented workflows: a CS session can only be initiated once a default LE asynchronous connection logical transport (LE ACL) has been established between the Initiator and Reflector, and its termination ends the CS session~\cite{b_core60}. For large-scale deployments, this requirement becomes a bottleneck. Practical System-on-Chip (SoC) implementations support on the order of 20 concurrent LE ACL connections, imposing a scalability ceiling per radio~\cite{b_nikodem2025}. A recent review of indoor localization technologies reports that BLE-based deployments support approximately seven tags per anchor in practice, with interference scaling with device count~\cite{b_tim2026}. Before measurements can begin, the standard procedure additionally executes a multistage initiation comprising CS Security Enable, capabilities exchange, configuration, and CS Start~\cite{b_core60,b_cs_overview}. In dynamic scenarios with frequent partner switching, this incurs substantial latency because discovery and connection setup for each new partner can take seconds and dominate the overall time budget~\cite{b_gunia2026}. Beyond such overhead, CS itself has been shown to be largely robust against external 2.4~GHz cross-technology interferers~\cite{b_sheikh2025}. CS-to-CS channel collisions among many parallel pairs, however, have been highlighted as a distinct stressor in dense deployments~\cite{b_wieme2025}.

Prior work has recognized that one-to-one ranging scales poorly in dense BLE networks. Zand \emph{et al.} proposed phase-based group ranging with one-to-many and many-to-many tone exchanges, anchored in connection events or in a connectionless advertising-based variant~\cite{b_zand2019wcnc,b_zand2019pimrc}. An earlier industry patent described exchanging ranging data after a connectionless advertising/scan handshake to avoid the signaling overhead of a full Bluetooth connection~\cite{b_wulff2022}. More recently, a commercial-hardware evaluation of Bluetooth 6.0 CS suggested a synchronized backbone network to reduce the burden of connection-based result transfer~\cite{b_wieme2025}. These contributions predate the Bluetooth 6.0 CS specification or remain at the architectural-sketch level. To the best of our knowledge, no prior work combines a standardized CS execution path with system-wide coordination that delivers deterministic role, channel, and slot assignment together with a compact data plane and an end-to-end evaluation. This work closes that gap.

The core idea is to decouple CS from a connection-based context by combining (i) the Host Controller Interface LE CS Test command, which can initiate CS procedures without an established LE ACL connection, and (ii) Periodic Advertising with Responses (PAwR), which provides time-structured bidirectional communication between an advertiser and synchronized devices. The resulting architecture shifts coordination, configuration alignment, and data aggregation responsibilities to the application layer, enabling connectionless operation at scale with dynamic partner reconfiguration and arbitrary device-to-device pairings rather than only the conventional tag-to-anchor pattern.

The main contributions of this paper are:
\begin{itemize}
\item A connectionless CS orchestration architecture based on LE CS Test and PAwR, centered around a \emph{Central Orchestrator} (CO) and a Gateway (GW), supporting dynamic partner reconfiguration and arbitrary device-to-device measurement topologies.
\item Deterministic coordination mechanisms for device indexing, role assignment, Deterministic Random Bit Generator (DRBG) state derivation, channel management, and response slot assignment.
\item A compact, configuration-aware data plane with on-device preprocessing and serialization for efficient response reporting.
\item An experimental evaluation of feasibility and collision-stress robustness, an energy comparison against a connected baseline under both standardized Ranging Data exchange modes, and a timing-based scalability analysis.
\end{itemize}

The paper is organized as follows. Section~\ref{sec:background} reviews the background on CS and PAwR. Section~\ref{sec:architecture} presents the connectionless CS architecture. Section~\ref{sec:evaluation} reports the experimental evaluation. Section~\ref{sec:conclusion} concludes and outlines future work.

\section{Background}
\label{sec:background}
This section provides the technical background for the architecture presented in Section~\ref{sec:architecture}, covering the CS procedure and the DRBG-based coordination state exposed by the LE CS Test command, as well as the PAwR subevent structure, timing parameters, and response slot mechanism.

\begin{figure*}[!t]
\centering
\includegraphics[width=0.73\textwidth]{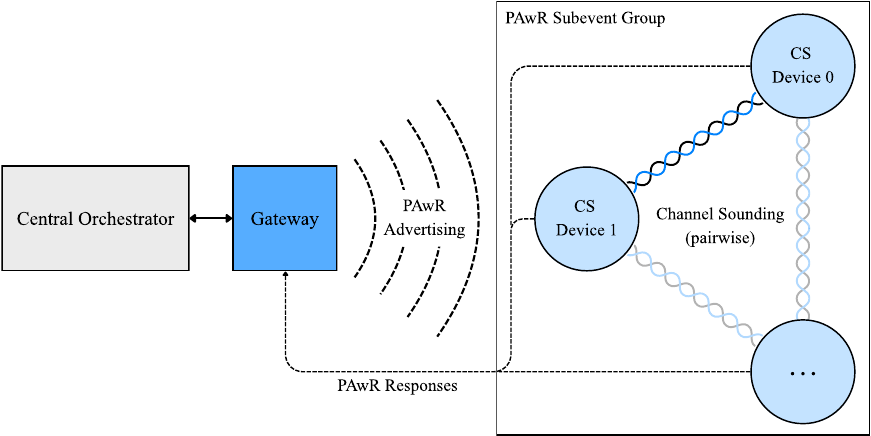}
\caption{High-level system architecture: The CO generates per-subevent measurement configurations and aggregates reports, the GW relays them over PAwR, and a synchronized CS device group executes the scheduled CS procedures.}
\label{fig:system_overview}
\end{figure*}

\subsection{Bluetooth Channel Sounding}
Bluetooth CS defines procedures between an Initiator and a Reflector. A CS procedure consists of one or more CS events, each comprising one or more CS subevents that in turn contain a sequence of CS steps, where the mode of each step determines the operation it performs~\cite{b_core60}. Each CS subevent begins with mode-0 steps for frequency-offset calibration. The Initiator measures the frequency offset of the Reflector's signal and uses it to compensate for frequency and timing errors in its transmissions during subsequent steps of that subevent~\cite{b_core60}. This paper focuses on PBR, using mode-2 steps for the subsequent CS tone exchanges. RTT (mode-1) and combined (mode-3) steps are out of scope.

CS measurements use a defined set of 72 channels in the 2.4~GHz band (excluding the primary advertising channels)~\cite{b_core60}. In this work, we use all 72 channels (1~MHz spacing) or a 37-channel subset (2~MHz spacing). The channel spacing $\Delta f$ sets the theoretical maximum one-way unambiguous range $d_{\mathrm{u}} = c / (2\,\Delta f)$, where $c$ is the speed of light and the factor $1/2$ accounts for the round-trip path, giving approximately 150~m at 1~MHz and approximately 75~m at 2~MHz. Each CS procedure uses up to four antenna paths, with the count denoted by $N_{\mathrm{AP}}$. The specific combination of Initiator and Reflector antennas is selected via an Antenna Configuration Index (ACI). The ACI-to-$N_{\mathrm{AP}}$ mapping is defined in the Bluetooth Core Specification~\cite{b_core60}. A more detailed description of the CS procedure is provided in~\cite{b_cs_overview}.

Each mode-2 step yields, per antenna path, a tone Phase Correction Term (PCT), a complex sample represented by in-phase and quadrature components, with phase (measured relative to the local oscillator) and magnitude that together describe the received signal. It also reports a per-path tone quality indication (high, medium, low, or unavailable) that flags the sample's reliability. A Reference Power Level (RPL) reported once per CS subevent provides the reference for converting the unitless tone PCT magnitude to absolute power. Beyond the per-path tones, each mode-2 step includes a CS tone extension slot. The tone quality indication also identifies whether an entry corresponds to this slot and whether a tone is expected to be present~\cite{b_core60}.

CS relies on a DRBG to derive per-step coordination values (e.g., access address updates and CS tone extension behavior). Since both Initiator and Reflector must instantiate the DRBG with identical state to derive matching per-step values, they must share the same DRBG configuration for a given CS procedure. Under the LE CS Test command, the standard security context establishment is omitted, so the DRBG key $K_{\mathrm{DRBG}}$ and the first 14~bytes of the counter nonce vector $V_{\mathrm{DRBG}}$ are initialized to all zeros~\cite{b_core60}. The 16-bit \texttt{DRBG\_Nonce} parameter, which populates bytes~14--15 of $V_{\mathrm{DRBG}}$, is therefore the only configurable field of the DRBG initialization state. The LE CS Test command additionally exposes an \texttt{Override\_Config} bitmask that allows the application to supply explicit deterministic values for selected DRBG-controlled behaviors, most notably the channel hopping sequence, which can be replaced by an application-provided channel list~\cite{b_core60}.

\subsection{Periodic Advertising with Responses}

PAwR extends periodic advertising to support bidirectional communication. Each PAwR event contains one or more subevents. A 16-bit event counter increments once per periodic advertising interval, wraps from \texttt{0xFFFF} to \texttt{0x0000}, and is reported to the application alongside each received subevent. Within a subevent, an advertiser transmits an \texttt{AUX\_SYNC\_SUBEVENT\_IND} Protocol Data Unit (PDU), after which synchronized devices transmit \texttt{AUX\_SYNC\_SUBEVENT\_RSP} PDUs in a set of response slots~\cite{b_core60}. Both PDUs use the Common Extended Advertising Payload Format and carry their host-provided application data in the \texttt{AdvData} field. In the proposed architecture, the respective \texttt{AdvData} contents are termed the \emph{configuration payload} and the \emph{response payload}.

The timing structure of a PAwR train is parameterized by the periodic advertising interval (time between consecutive PAwR events), the subevent interval, the response slot delay, and the response slot spacing, with up to 128 subevents per PAwR event and up to 255 response slots per subevent~\cite{b_core60}. A more detailed description of PAwR is provided in~\cite{b_pawr_guide}.

\section{System Architecture}
\label{sec:architecture}

This section presents the architecture of the proposed connectionless CS system. To decouple ranging from persistent LE ACL connections, the architecture orchestrates CS procedures initiated via the LE CS Test command using PAwR as a scalable communication backbone. The following subsections first derive the design requirements from the responsibilities transferred by the LE CS Test command, then introduce the system components and their end-to-end interaction, and finally detail the control plane, deterministic coordination mechanisms, and compact data plane that realize these requirements.

\subsection{Design Drivers and Requirements}
In the standard procedure, CS requires an established LE ACL connection between the Initiator and Reflector and then follows a multistage initiation sequence before measurements can begin. This sequence comprises CS Security Enable, which derives the cryptographic keys for the DRBG; CS Capabilities Exchange, where the Initiator-Reflector pair negotiates supported features; CS Configuration, which aligns measurement parameters; and CS Start, which establishes temporal synchronization anchored to the periodic connection events of the underlying LE ACL connection~\cite{b_core60}. Once initiated, the Bluetooth stack automates role coordination, security context management, channel sequencing via the DRBG, and timing. Measurement results are subsequently transferred over the same LE ACL connection using the Generic Attribute Profile (GATT)-based Ranging Service (RAS) and Ranging Profile (RAP)~\cite{b_ras,b_rap}.

The LE CS Test command bypasses this entire initiation procedure, providing direct access to a single CS procedure without requiring an LE ACL connection~\cite{b_core60}. However, this shifts all coordination responsibilities (configuration alignment, role assignment, synchronization, DRBG state management, and result aggregation) from the automated stack to the application layer. Based on these transferred responsibilities, the design is driven by the following requirements:
\begin{itemize}
\item \textbf{Coordination and synchronization:} use PAwR subevent reception as a high-precision timing anchor, assign persistent device identifiers, and deterministically map devices to noncolliding response slots.
\item \textbf{Configuration and DRBG state management:} deterministically assign Initiator/Reflector roles, align CS parameters per pair, manage nonoverlapping channel sequences, and derive \texttt{DRBG\_Nonce} values that are distinct across all CS procedures within the same subevent group and update cycle and vary across successive cycles.
\item \textbf{Data management and transmission:} preprocess and compact raw results, define a serialization format that can be contextually parsed, and correlate delayed PAwR responses with the corresponding measurement configuration.
\item \textbf{Platform constraints:} respect hardware-imposed response slot timing limits and Controller buffer constraints, and avoid consecutive-slot transmissions where unsupported by the platform.
\end{itemize}

\begin{figure}[!t]
\centering
\includegraphics[width=\columnwidth]{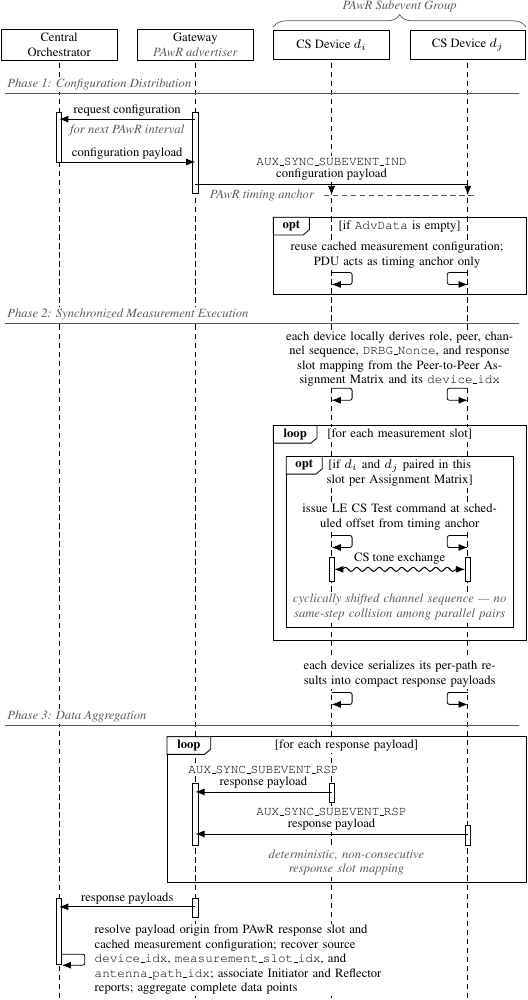}
\caption{One complete update cycle between the CO, the GW, and an example CS device pair $(d_i, d_j)$ within the synchronized subevent group. Only one paired exchange is shown for clarity, while in general each device executes the same flow whenever the Peer-to-Peer Assignment Matrix schedules it. The three phases comprise configuration distribution via the \texttt{\bfseries AUX\_SYNC\_SUBEVENT\_IND} PDU (which also serves as the PAwR timing anchor), synchronized CS tone exchange between paired devices, and deterministic response slot transmission with pair aggregation at the CO.}
\label{fig:update_cycle}
\end{figure}

\subsection{System Overview}
\label{sec:system_overview}
To address these requirements, the architecture employs a three-tier design (see Fig.~\ref{fig:system_overview}). Battery-powered \textbf{CS devices} execute CS procedures, preprocess results, and transmit compact reports via PAwR responses. Each device can act as either Initiator or Reflector, with the role derived deterministically for each measurement pair. Although localization deployments typically use devices with known positions as anchors and devices to be localized as tags, the architecture is not limited to the conventional tag-to-anchor pattern and also supports arbitrary device-to-device pairings, including tag-to-tag measurements. A \textbf{GW} acts as the PAwR advertiser, relaying configuration payloads from the CO to the synchronized CS devices and forwarding response payloads upstream. The \textbf{CO} generates per-subevent measurement configurations, maintains measurement context across PAwR intervals, and aggregates Initiator and Reflector reports into paired data points for downstream processing.

The following paragraphs trace one complete update cycle to illustrate how the three components interact (see Fig.~\ref{fig:update_cycle}).

\textbf{Phase~1: Configuration Distribution.}
The GW requests the measurement configuration for an upcoming PAwR subevent from the CO. The CO responds with a configuration payload containing a bit-packed configuration byte, the number of measurement slots per cycle, and a \emph{Peer-to-Peer Assignment Matrix} that specifies which device pairs perform measurements in each slot. The GW advertises this configuration payload in the corresponding PAwR subevent to all synchronized devices.

\textbf{Phase~2: Synchronized Measurement Execution.}
Upon reception, each device deserializes the payload and uses its \texttt{device\_idx} together with the Assignment Matrix to derive its role (Initiator or Reflector) and its peer for each measurement slot. At the scheduled time offset after subevent reception, both peers issue the LE CS Test command simultaneously, executing the CS tone exchange without an LE ACL connection. This process repeats for each configured measurement slot within the cycle.

\textbf{Phase~3: Data Aggregation.}
After completing all CS procedures, each device serializes its results into compact payloads and transmits them in deterministically assigned PAwR response slots. The GW forwards these payloads to the CO, which pairs the Initiator and Reflector reports into complete data points.

\subsection{Control Plane via PAwR}
\label{sec:control_plane}
The communication backbone underlying the update cycle described above is PAwR, with the GW as advertiser bridging the wireless CS devices and the CO. The GW is intentionally kept simple, without measurement-specific logic, and handles the timing-critical wireless exchange.

\textbf{Timing anchor.}
Devices first synchronize to the PAwR train by scanning for the advertiser and then remain locked to the periodic subevent timing. Each device treats reception of the \texttt{AUX\_SYNC\_SUBEVENT\_IND} PDU as the common timing reference for a deterministic schedule. The schedule triggers one or more LE CS Test commands at computed offsets such that both peers open their measurement windows synchronously.

\textbf{Predictive provisioning and context management.}
PAwR requires that subevent data are provisioned ahead of transmission. In the reference firmware stack, the GW receives the request for a given subevent one full periodic interval before that subevent is sent and therefore requests the corresponding measurement configuration from the CO in advance. Once distributed, a measurement configuration remains valid until a new configuration payload is advertised. If a subsequent \texttt{AUX\_SYNC\_SUBEVENT\_IND} carries no \texttt{AdvData}, devices reuse the previously received configuration payload and treat the PDU only as the timing anchor for the next cycle. The CO caches the transmitted measurement configuration for one full interval so that responses arriving in the next interval can be parsed against the correct context.

\textbf{Response handling constraints.}
On the evaluated platform, transmitting in consecutive response slots can fail when slot spacing is small and slot utilization is high. The response mapping therefore avoids consecutive-slot transmissions per device. In addition, bursts of incoming PAwR responses may exceed the Host processing rate. Sufficiently large Controller report buffers are configured to absorb bursts until the Host drains them.

\subsection{Deterministic Coordination Mechanisms}
\label{sec:coordination}
Within this PAwR-based framework, each device combines the cycle-wide shared state distributed via PAwR (most notably the Peer-to-Peer Assignment Matrix) with its persistent \texttt{device\_idx} and the current \texttt{measurement\_slot\_idx}. It uses these inputs to derive its pair assignment, role, DRBG state, channel sequence, and response slot assignment deterministically. The \texttt{antenna\_path\_idx} is additionally used when mapping per-path results to PAwR response slots. Because each device reconstructs its coordination state locally, the CO can change measurement partners between update cycles by simply distributing an updated Assignment Matrix, with no pairwise negotiation. In the current prototype, dynamic reconfiguration is limited to partner switching within the same PAwR subevent group. Cross-group reassignment remains future work.

\textbf{Assignment Matrix and device activity.}
Pair assignments for an update cycle are specified by a Peer-to-Peer Assignment Matrix of size $N_{\mathrm{dev}} \times N_{\mathrm{ms}}$, where $N_{\mathrm{dev}}$ is the number of CS devices in the subevent group and $N_{\mathrm{ms}}$ is the number of configured measurement slots per cycle. Each entry $(\texttt{device\_idx}, \texttt{measurement\_slot\_idx})$ maps either to a peer \texttt{device\_idx} (specifying a measurement pair for that slot) or to a reserved no-peer value indicating that the device performs no measurement in that slot. A device is termed \emph{active in a slot} if its matrix entry for that slot is a peer assignment, and \emph{active in the cycle} if it is active in at least one slot. The per-cycle CS-procedure count of a device, \mbox{$N_{\mathrm{meas}} \in \{0, \ldots, N_{\mathrm{ms}}\}$}, equals the number of slots in which it is active. $N_{\mathrm{ms}}$ is thus the scheduling capacity per device, whereas $N_{\mathrm{meas}}$ is the actual number of CS procedures the device executes.

\textbf{Role assignment and DRBG state derivation.}
For each measurement pair, the device with the lower \texttt{device\_idx} acts as Initiator and the device with the higher index acts as Reflector, so both devices independently derive the same role assignment. The Initiator's device index is also used to derive the DRBG state. As established in Section~\ref{sec:background}, under the LE CS Test command \texttt{DRBG\_Nonce} is the only configurable field of the DRBG initialization state. It is derived deterministically as
\begin{equation}
n_{\mathrm{drbg}} = (c_{\mathrm{pawr}} + i_{\mathrm{ini}} \cdot N_{\mathrm{ms}} + i_{\mathrm{ms}}) \bmod 2^{16},
\label{eq:drbg_nonce}
\end{equation}
where $n_{\mathrm{drbg}}$ is the 16-bit value supplied to the \texttt{DRBG\_Nonce} parameter of the LE CS Test command, $c_{\mathrm{pawr}}$ is the PAwR event counter, \mbox{$i_{\mathrm{ini}} \in \{0, \ldots, N_{\mathrm{dev}}-1\}$} is the \texttt{device\_idx} of the Initiator within a subevent group of $N_{\mathrm{dev}}$ devices, \mbox{$i_{\mathrm{ms}} \in \{0, \ldots, N_{\mathrm{ms}}-1\}$} is the \texttt{measurement\_slot\_idx}, and $N_{\mathrm{ms}}$ is the number of measurement slots per cycle. As long as $N_{\mathrm{dev}} \cdot N_{\mathrm{ms}} \leq 2^{16}$, the mapping $(i_{\mathrm{ini}}, i_{\mathrm{ms}}) \mapsto i_{\mathrm{ini}} \cdot N_{\mathrm{ms}} + i_{\mathrm{ms}}$ is injective, so every CS procedure within a subevent group and update cycle receives a distinct nonce. For fixed $(i_{\mathrm{ini}}, i_{\mathrm{ms}})$, the nonce traverses all $2^{16}$ values as $c_{\mathrm{pawr}}$ advances. With $K_{\mathrm{DRBG}}=0$ (see Section~\ref{sec:background}), the \texttt{DRBG\_Nonce} derivation serves coordination rather than security: distinct nonces within a subevent group and update cycle prevent parallel CS procedures from sharing a DRBG state. An optional application-layer security extension remains future work (see Section~\ref{sec:conclusion}).

\textbf{Deterministic channel management.}
To prevent interference between parallel measurements, each pair receives an application-provided channel sequence via the \texttt{Override\_Config} channel-sequence override of the LE CS Test command (see Section~\ref{sec:background}). The sequence is constructed as a cyclic shift of a base CS channel list $\mathcal{C} = (c_0, c_1, \ldots, c_{N-1})$. In each measurement slot $i_{\mathrm{ms}}$, the pairs active in that slot are ordered deterministically (ascending by Initiator index, which is unique per pair within a slot) and assigned positions \mbox{$i_{\mathrm{pair}} \in \{0, \ldots, P_{\mathrm{ms}}-1\}$}, where $P_{\mathrm{ms}}$ is the number of active pairs in that slot. Both partners derive $i_{\mathrm{pair}}$ identically from the shared Peer-to-Peer Assignment Matrix. The channel used at CS step $i_{\mathrm{step}}$ is $c_j$ with
\begin{equation}
j = (i_{\mathrm{step}} + i_{\mathrm{pair}} + i_{\mathrm{ms}} + c_{\mathrm{pawr}}) \bmod N,
\label{eq:channel_shift}
\end{equation}
where $N$ is the number of channels in the base channel list $\mathcal{C}$ and, for the single-pass configuration considered here, \mbox{$i_{\mathrm{step}} \in \{0, \ldots, N-1\}$} indexes the mode-2 steps of the CS procedure. Here $N=72$ for the 1~MHz configuration and $N=37$ for the 2~MHz subset (see Section~\ref{sec:background}). Within one measurement slot, pairs with distinct $i_{\mathrm{pair}}$ use different channels at every step, since their shifts differ by construction and the modulo-$N$ operation preserves this distinction as long as $P_{\mathrm{ms}} \leq N$. Up to $N$ Initiator-Reflector pairs can therefore be scheduled in parallel without same-step channel collision, regardless of pairing strategy. This supports subevent groups of up to $2N$ devices with arbitrary, dynamically reconfigurable pairings. The $i_{\mathrm{ms}}$ and $c_{\mathrm{pawr}}$ terms additionally rotate each pair's channel sequence across measurement slots and update cycles, changing its step-to-channel assignment without creating same-step collisions. A shared pseudorandom permutation of $\mathcal{C}$ could likewise be generated for each measurement slot and update cycle before the pair-specific shifts. It would preserve this collision-avoidance property while varying the temporal channel order, which could reduce repeated overlap with time-varying external interference.

\textbf{Deterministic response slot assignment.}
Each response payload occupies a distinct PAwR response slot and carries the per-path result of one (\texttt{measurement\_slot\_idx}, \texttt{antenna\_path\_idx}) combination. At 2~MHz channel spacing, where the per-path payload is small enough, two such results are concatenated into one response payload (detailed in Section~\ref{sec:data_plane}). With $N_{\mathrm{AP}}$ antenna paths per CS procedure, each active device therefore transmits up to
\begin{equation}
N_{\mathrm{rsp}} = \left\lceil \frac{N_{\mathrm{ms}}\,N_{\mathrm{AP}}}{k} \right\rceil
\label{eq:n_rsp}
\end{equation}
response payloads per cycle, where $k$ is the number of per-path results concatenated per response payload ($k=1$ at 1~MHz spacing, $k=2$ at 2~MHz spacing).

In each update cycle, the active devices are ordered by increasing \texttt{device\_idx} and assigned positions \mbox{$i_{\mathrm{act}} \in \{0, \ldots, N_{\mathrm{act}}-1\}$}, where $N_{\mathrm{act}}$ is the number of active devices. For slot allocation, active devices are grouped into adjacent pairs by this position (positions $2a$ and $2a{+}1$ form one allocation group, independently of measurement pairing). Each group is allocated a contiguous block of $N_{\mathrm{blk}}$ response slots, and within that block the two devices interleave on even and odd slot offsets. An active device at position $i_{\mathrm{act}}$ transmits its $i_{\mathrm{rsp}}$-th response payload of the cycle (\mbox{$i_{\mathrm{rsp}} \in \{0, \ldots, N_{\mathrm{rsp}}-1\}$}) in slot $i_{\mathrm{rs}}$ with
\begin{equation}
i_{\mathrm{rs}} = \left\lfloor \frac{i_{\mathrm{act}}}{2} \right\rfloor \cdot N_{\mathrm{blk}} + 2\,i_{\mathrm{rsp}} + (i_{\mathrm{act}} \bmod 2),
\label{eq:response_slot}
\end{equation}
where $N_{\mathrm{blk}}$ is the per-group block size, $N_{\mathrm{rs}}$ is the configured number of PAwR response slots per subevent, and \mbox{$i_{\mathrm{rs}} \in \{0, \ldots, N_{\mathrm{rs}}-1\}$} is the resulting PAwR response slot index. Different allocation groups occupy disjoint blocks, and within a group the parity term separates the two devices, so the mapping is injective. The allocation is valid under the sizing constraints $N_{\mathrm{blk}} \geq 2\,N_{\mathrm{rsp}}$ (block large enough for all interleaved payloads) and $\lceil N_{\mathrm{act}}/2 \rceil \cdot N_{\mathrm{blk}} \leq N_{\mathrm{rs}}$ (overall allocation fits within the configured response slot count). Because successive values of $i_{\mathrm{rsp}}$ advance the slot index by two, no device ever occupies consecutive slots, respecting the platform constraint noted in Section~\ref{sec:control_plane}. Since $N_{\mathrm{rs}}$ is fixed at PAwR configuration time, restricting the allocation to active devices rather than to all devices in the subevent group allows the CO to treat the response slot window as a budget balanced between $N_{\mathrm{act}}$ and $N_{\mathrm{blk}}$: large subevent groups with few devices active per cycle, or configurations with different antenna path counts and 1~MHz or 2~MHz channel spacing, can be accommodated within the same fixed response slot count.

\subsection{Compact Data Plane}
\label{sec:data_plane}
With coordination mechanisms in place, the remaining architectural component is efficient data handling. In a standard connected procedure, CS results are transferred using GATT-based RAS and RAP~\cite{b_ras,b_rap}. In the connectionless architecture, this path is unavailable. Instead, each CS device must preprocess, compact, and transmit its results as PAwR response payloads to the CO.

\textbf{On-device preprocessing and compact serialization.}
Each CS device first interprets the tone quality indication to correctly handle measurements from CS tone extension slots. PCTs originating from the same channel and antenna path can be averaged to improve the Signal-to-Noise Ratio (SNR). Mode-0 step data, which the Controller has already used internally for frequency-offset calibration, are discarded. The remaining mode-2 step data are then sorted in a predefined order (ascending by channel index and antenna path) so that step channel indices and antenna-permutation fields can be omitted: their information is implicitly encoded in the immutable ordering. For each mode-2 step, the payload is reduced to the 3-byte tone PCT and a 2-bit quality indication representing the four quality levels (high, medium, low, unavailable). The entire \emph{per-path result} then consists of a single RPL byte followed by the stream of sorted and compacted mode-2 step data. The RPL is included in each per-path result rather than reported once per subevent as in the standard RAS Subevent Header~\cite{b_ras}, so that each per-path result remains independently parsable and the loss of a response slot does not invalidate the remaining paths of the same CS procedure.

In the standard procedure, assuming a single antenna path and 3 mode-0 steps, the Ranging Data Body for a CS subevent amounts to at least 744~bytes with 72 mode-2 steps (1~MHz spacing), or 394~bytes with 37 mode-2 steps (2~MHz spacing). The 1~MHz case already exceeds the maximum attribute value length of 512~octets~\cite{b_core60} and must therefore be split across multiple Ranging Data segments, each carrying at most $(\mathrm{ATT\_MTU} - 4)$~octets, where $\mathrm{ATT\_MTU}$ denotes the Attribute Protocol Maximum Transmission Unit~\cite{b_ras}. Each transmitted segment is prefixed by a 1-byte Segmentation Header. The serialization comparison excludes this byte and the lower-layer protocol overhead on both paths. The optimized payload reduces the body sizes to 235~bytes (1~MHz spacing) and 122~bytes (2~MHz spacing), representing reductions of approximately 68\% and 69\%, respectively. This reduction is what makes it practical to report a complete CS result inside PAwR response slots rather than through a segmented connection-based transfer.

\textbf{Configuration and response payload semantics.}
The configuration payload conveys only the state required to deterministically reproduce the measurement schedule: a bit-packed configuration byte (standby mode, channel spacing, ACI), the number of measurement slots per cycle, and the Peer-to-Peer Assignment Matrix as described in Section~\ref{sec:coordination}. The slot count is transmitted explicitly so that the matrix dimensions can be recovered from the serialized representation, whose specific form (e.g., a dense two-dimensional array or, for larger groups, a compact encoding) does not affect the coordination derivations.

The response payload carries the compact per-path result described above, whose interpretation depends on the shared measurement configuration. Because the CO generates it and the devices receive it via PAwR, both sides hold matching channel and antenna path context, which need not be repeated in the response payload.

When response payloads arrive, the CO resolves the origin of each payload using the PAwR response slot index together with the cached measurement configuration, recovering the source \texttt{device\_idx}, \texttt{measurement\_slot\_idx}, and \texttt{antenna\_path\_idx}. It then associates the Initiator and Reflector reports for each pair and aggregates them into complete data points that can be passed to downstream processing stages such as distance estimation.

\textbf{Payload limits and timing implications.}
The response slot spacing must accommodate the actual on-air packet duration plus platform processing overhead. For the evaluated prototype, a spacing of 1.25~ms using the LE 2M Physical Layer (PHY) is sufficient to reliably transmit a concatenated 244-byte payload in the 2~MHz spacing mode. This confirms that the compact serialization keeps reporting within practical PAwR timing limits.

\section{Experimental Evaluation}
\label{sec:evaluation}
To assess the practical viability of the proposed architecture, this section presents a proof-of-concept evaluation structured around a direct comparison with the standard connection-based CS procedure. The evaluation covers three dimensions: collision robustness under simulated dense-deployment conditions, energy efficiency, and system scalability and capacity. The methodology below introduces the shared prototype platform, common CS settings, proof-of-concept processing pipeline, and metrics.

\newpage
\subsection{Methodology}
\subsubsection{Prototype Platform}
\begin{figure}[!t]
\centering
\includegraphics[width=\columnwidth]{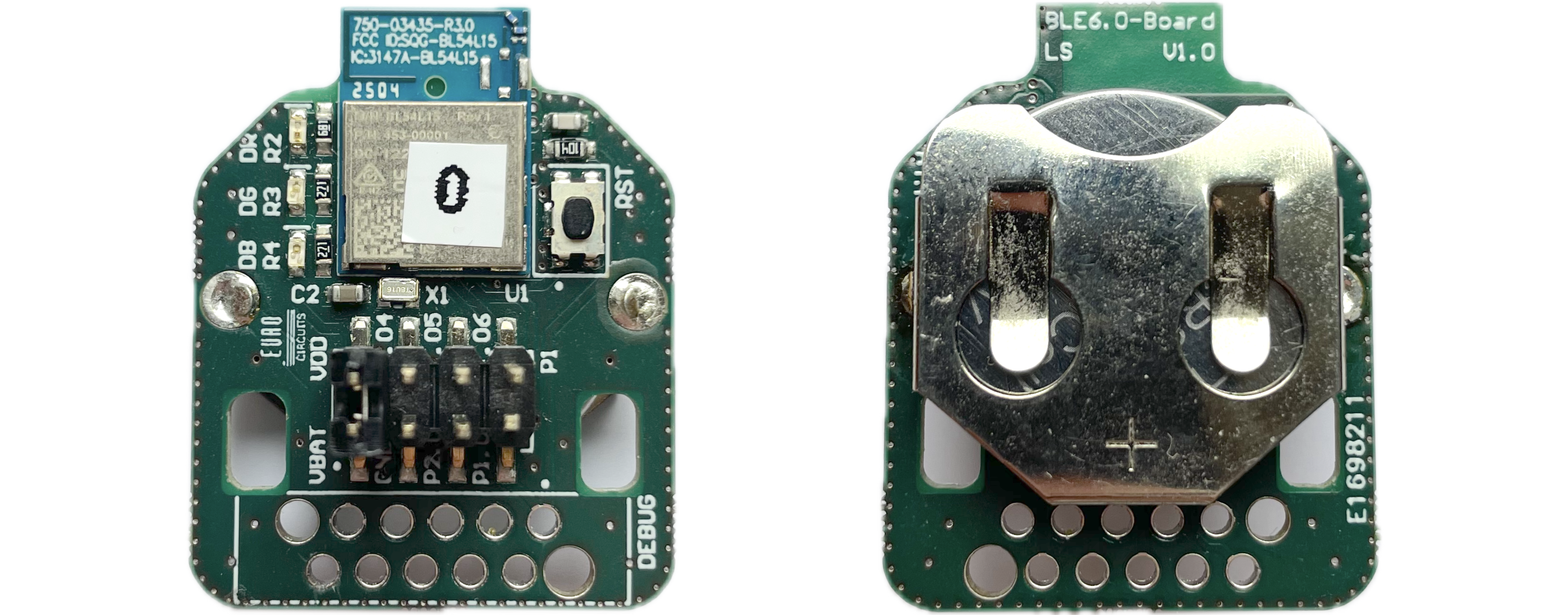}\\[2pt]
\makebox[0.5\columnwidth]{\hspace{5mm}\footnotesize (a) Top view}%
\makebox[0.5\columnwidth]{\footnotesize (b) Bottom view\hspace{5mm}}
\caption{Custom CS device based on the Ezurio BL54L15 module: (a)~top view with SoC module, LFXO, and supporting peripherals; (b)~bottom view with the CR2032 coin cell in its holder.}
\label{fig:prototype_platform}
\end{figure}

The prototype comprises eight identical custom CS devices based on the Ezurio BL54L15 module, which integrates the Nordic Semiconductor nRF54L15 SoC with native support for both CS and PAwR. Each device is powered by a CR2032 coin cell (3~V, 240~mAh) and equipped with a dedicated 32.768~kHz Low-Frequency Crystal Oscillator (LFXO) that serves as the clock source for the Global Real-Time Counter (GRTC) system timer, enabling low-power timekeeping with the tickless kernel. The firmware is built with nRF Connect SDK v3.1.0 on the Zephyr real-time operating system and uses the Nordic SoftDevice Controller. An nRF54L15 Development Kit serves as the GW, connected to the CO via a UART-to-USB link. The CO runs a Python-based multithreaded pipeline for measurement configuration generation, data aggregation, and distance estimation. The custom CS device is shown in Fig.~\ref{fig:prototype_platform}.

\subsubsection{Configurations}
Unless stated otherwise, each CS procedure consists of one CS subevent with 3 mode-0 steps followed by either 72 mode-2 steps (1~MHz spacing) or 37 mode-2 steps (2~MHz spacing). We refer to these baseline scenarios as 1$\times$72~ch.\ and 1$\times$37~ch., where the leading factor denotes the number of CS procedures executed per update cycle. The nominal update cycle is 1~s unless stated otherwise. Within each Initiator-Reflector pair, both devices use the same CS transmit power. Experiment-specific workload parameters are given in the respective subsections. The Appendix lists the shared radio, controller, and system settings and the additional parameters specific to the connected baseline.

\subsubsection{Processing Pipeline (Proof-of-Concept)}
\label{sec:pipeline}
Distance estimation uses a proof-of-concept pipeline based on the Inverse Fast Fourier Transform (IFFT). For each CS procedure, the Initiator and Reflector PCTs are combined via complex multiplication to cancel the unknown local oscillator starting phases~\cite{b_cs_overview}. Spectral gaps at channels excluded from the CS channel set (e.g., primary advertising channels) are filled with linear interpolation, and the resulting frequency-domain vector is zero-padded to 2048 entries before applying an IFFT to obtain the Channel Impulse Response (CIR). The earliest local CIR magnitude peak exceeding 40\% of the global maximum is selected, with the global maximum used if no earlier peak qualifies. A heuristic based on the preceding flank corrects the peak position before conversion to distance, with parabolic interpolation applied when the corrected sample is a local maximum. A static calibration offset is subtracted to compensate for hardware-related delays (e.g., antenna group delay). This estimator is intentionally simple and is used to confirm measurability and to quantify the impact of collision stress. It is \emph{not} presented as an accuracy benchmark. Advanced estimators and multipath/non-line-of-sight optimization are out of scope.

\subsubsection{Metrics}
For the collision study, we report Mean Absolute Error (MAE), peak absolute error, 90th-percentile absolute error ($P_{90}$), and Standard Deviation (STD) of the signed error (estimated minus real distance). The metrics are computed separately for each pair over all measured distances. Reported table values are rounded to at most two decimal places, while all derived quantities are calculated from unrounded values.

\subsection{Channel Collision Robustness in Dense Deployments}
\label{sec:collision}
\subsubsection{Setup}
\begin{figure}[!t]
\centering
\includegraphics[width=\columnwidth]{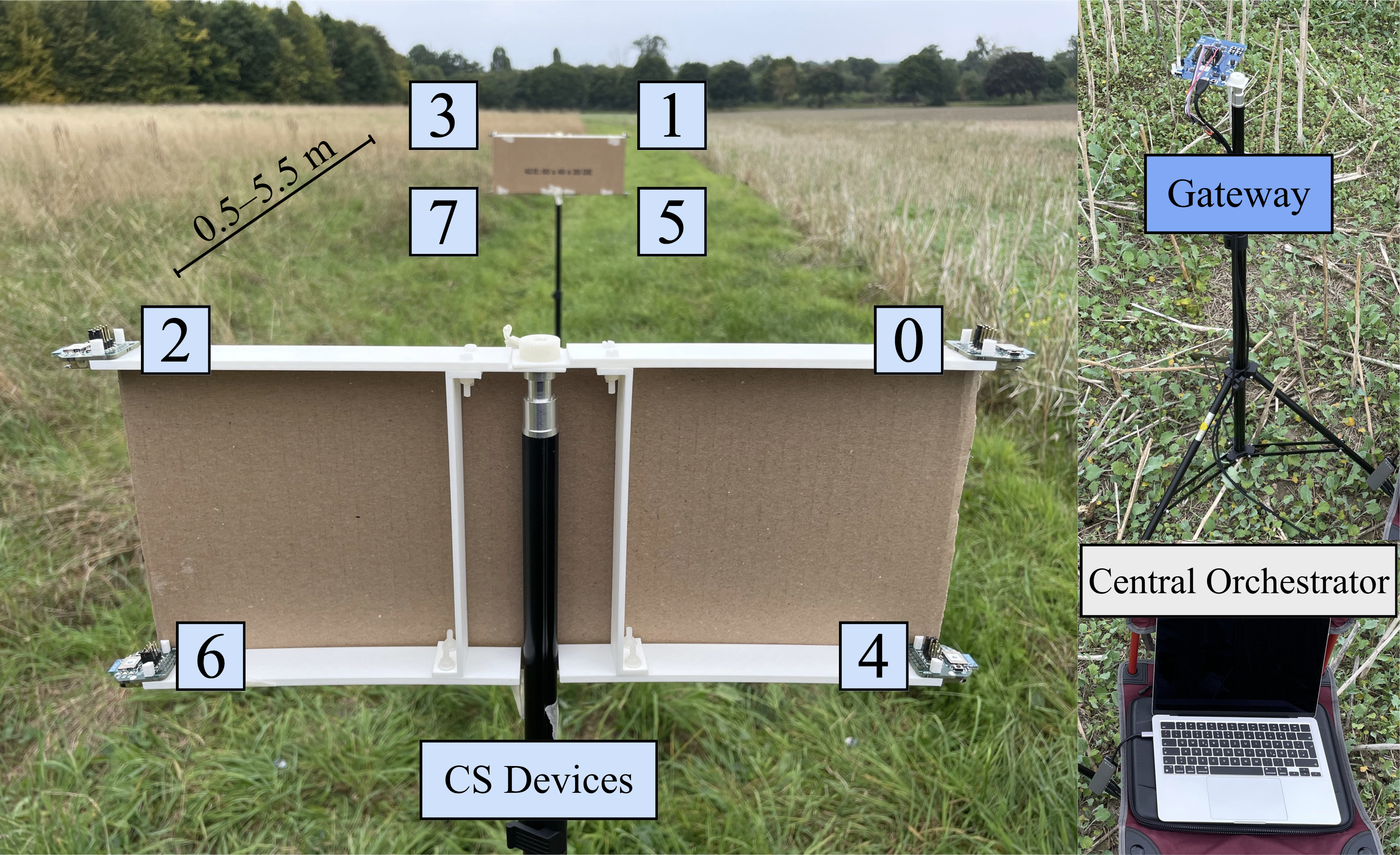}\\[2pt]
\makebox[0.78\columnwidth]{\footnotesize (a)}%
\makebox[0.22\columnwidth]{\footnotesize (b)}
\caption{Distance measurement setup. (a)~Eight CS devices form four ranging pairs (0$\leftrightarrow$1, 2$\leftrightarrow$3, 4$\leftrightarrow$5, 6$\leftrightarrow$7) across two tripods. The two devices of each pair sit on opposite tripods, separated by the measured range (varied over 0.5--5.5~m). The collision-free pairs (0$\leftrightarrow$1, 2$\leftrightarrow$3) share the upper mount (156~cm height), and the collision-stress pairs (4$\leftrightarrow$5, 6$\leftrightarrow$7) the lower mount (140~cm height). On each mount, the two pairs are offset 41~cm horizontally. (b)~GW (nRF54L15 Development Kit on tripod) and CO (laptop).}
\label{fig:distance_setup}
\end{figure}

Measurements were performed in an open field with a grass surface to minimize multipath reflections. All eight CS device prototypes were mounted pairwise on two tripods (see Fig.~\ref{fig:distance_setup}). Devices within each pair were aligned parallel for consistent antenna polarization. For this collision study, CS transmit (TX) power was set to $+8$~dBm to maximize the SNR.

Device pairs 0$\leftrightarrow$1 and 2$\leftrightarrow$3 operated under the proposed system with collision-free channel assignments (upper mounts). Pairs 4$\leftrightarrow$5 and 6$\leftrightarrow$7 simulated collision-stress operation via fixed channel overlaps (lower mounts, see Section~\ref{sec:collision_model}). To limit temporal variation in the experimental conditions, both scenarios were executed within the same update cycle and anchored to the same PAwR subevent. The collision-free pairs first executed four consecutive CS procedures in parallel, immediately followed by the collision-stress pairs executing the same sequence. Distances ranged from 0.5 to 5.5~m in 0.5~m increments, measured with a laser distance meter. At each distance, a single recording covered 30 or 31 update cycles. After excluding procedures without a complete result, 117--124 complete CS results remained per distance and pair. The collision-free pairs also served as calibration reference: the static offset subtracted by the distance estimator (see Section~\ref{sec:pipeline}) was determined as the mean error of the uncorrected estimates over these two pairs, yielding an offset of 1.24~m.

Since the deterministic channel management of the proposed architecture (see Section~\ref{sec:coordination}) prevents channel overlaps by design, the collision-free results with two simultaneously measuring pairs represent the expected behavior with respect to same-step channel collisions for up to 72 simultaneous pairs. Aggregate effects of many concurrent transmitters (e.g., adjacent-channel interference) are not captured by this setup.

\subsubsection{Interference-Collision Model} 
\label{sec:collision_model}
In standard CS operation, each Initiator-Reflector pair uses a randomized channel sequence. When $P$ pairs measure simultaneously over $N$ channels, the per-step collision probability $p$ (i.e., the probability that at least one of the other $P-1$ pairs uses the same channel) is
\begin{equation}
p = 1 - \left(\frac{N-1}{N}\right)^{P-1}.
\label{eq:collision_prob}
\end{equation}
By linearity of expectation,
\begin{equation}
\mathrm{E}[X] = N \cdot \left[1 - \left(\frac{N-1}{N}\right)^{P-1}\right].
\label{eq:expected_overlaps}
\end{equation}
The random variable $X$ counts the mode-2 steps in one CS procedure at which the considered pair shares its channel with at least one other pair.

Preliminary tests showed that overlap counts beyond approximately 24 caused some CS procedures to fail entirely. We therefore chose $P=30$ simultaneous pairs with $N=72$ channels, for which~\eqref{eq:expected_overlaps} yields $\mathrm{E}[X]\approx 24$ overlapping channels per procedure, representing the highest collision load that still permits reliable CS execution.

To stress collision robustness without requiring 30 physical pairs, the collision-stress pairs (4$\leftrightarrow$5, 6$\leftrightarrow$7) ran the proposed system firmware with its deterministic channel management overridden by a channel sequence containing 24 randomly positioned overlaps, matching the expected collision count for $P=30$.
\subsubsection{Results and Analysis}
\begin{figure}[!t]
\centering
\includegraphics[width=\columnwidth]{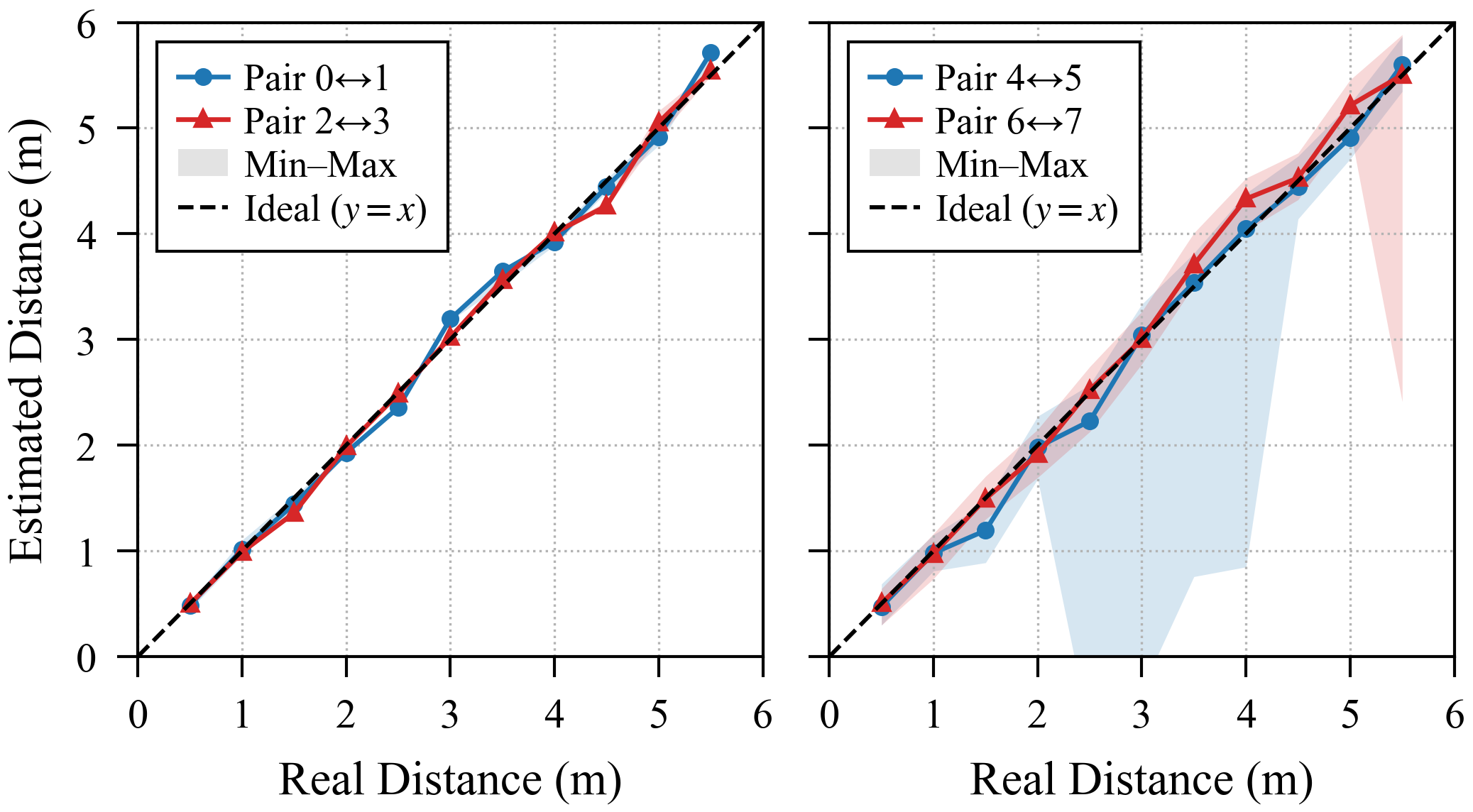}\\[2pt]
\makebox[0.5\columnwidth]{\footnotesize (a) Collision-free}%
\makebox[0.5\columnwidth]{\footnotesize (b) Collision-stress}
\caption{Estimated versus real distance for 1~MHz channel spacing (72 channels). Solid lines denote the per-pair mean, shaded bands the min--max range, and the dashed line the ideal $y=x$. (a)~Collision-free operation (pairs 0$\leftrightarrow$1, 2$\leftrightarrow$3): the min--max band is so narrow that it is barely visible. (b)~Operation with simulated channel collisions (pairs 4$\leftrightarrow$5, 6$\leftrightarrow$7, 24 overlapping channels): substantially wider band with large outliers. The min--max band in (b) extends below the displayed lower limit of 0~m.}
\label{fig:collision_results}
\end{figure}

Fig.~\ref{fig:collision_results} and Table~\ref{tab:collision_table} summarize the proof-of-concept results for 1~MHz channel spacing (72 channels). The collision-free pairs serve as the stable baseline. Under collision stress, peak errors increase from 25 to over 300~cm and STD more than doubles.

These results confirm that deterministic channel management effectively eliminates collision-induced outliers and yields stable, evaluable measurements even with multiple pairs measuring simultaneously.

\begin{table}[!t]
\caption{Proof-of-concept ranging performance with and without simulated channel collisions (1~MHz channel spacing, 72 channels)}
\label{tab:collision_table}
\centering
\begin{tabular*}{\columnwidth}{@{\hskip\tabcolsep\extracolsep{\fill}}lrrrr@{\hskip\tabcolsep}}
\toprule
\multirow{2}{*}[-0.6ex]{\textbf{Scenario}} & \multicolumn{4}{c}{\textbf{Error (cm)}} \\
\cmidrule(lr){2-5}
& \textbf{MAE} & \textbf{Peak} & $\boldsymbol{P_{90}}$ & \textbf{STD} \\
\midrule
{\raggedright Collision-free (0$\leftrightarrow$1)} & 10 & 25 & 20 & 12 \\
{\raggedright Collision-free (2$\leftrightarrow$3)} & 6 & 25 & 16 & 9 \\
{\raggedright Collision-stress (4$\leftrightarrow$5)} & 14 & 331 & 27 & 29 \\
{\raggedright Collision-stress (6$\leftrightarrow$7)} & 13 & 309 & 28 & 20 \\
\bottomrule
\end{tabular*}
\end{table}

\subsection{Energy Efficiency}
\label{sec:energy}
\subsubsection{Setup}
Current draw was measured using the Nordic Power Profiler Kit II (PPK2) at a sampling rate of 100~kS/s. The PPK2 was configured as a source meter with a supply voltage of 3~V. The connected baseline and proposed system were measured with identical CS air-interface parameters, whose experiment-specific values are listed in the Appendix. We report the consumed charge $Q$ over an integration interval of duration $T_{\mathrm{int}}$ as
\begin{equation}
Q = \int_0^{T_{\mathrm{int}}} i(t)\,\mathrm{d}t,
\end{equation} 
where $i(t)$ denotes the instantaneous supply current. We express $Q$ in microcoulombs. Under a fixed supply voltage $V$, the corresponding energy is $E=VQ$, so comparisons in $Q$ directly reflect relative energy consumption. For the energy measurements, we report charge per update cycle with $T_{\mathrm{upd}}\approx1~\mathrm{s}$. For each active contribution (e.g., \emph{CS} and \emph{Data TX}), we integrate the current above the local sleep baseline over the corresponding subinterval. The local sleep current for each evaluated cycle was estimated as the median of all samples in nonoverlapping 1~ms inactive windows.

Each energy run measured one device in one CS role with one partner. The four-partner scenarios are component-based projections. For each system, steady-state results are based on three independent averages of 20 update cycles: separate connections for the connected baseline and separately started runs for the proposed system. Means and sample standard deviations were computed across these three averages. Each initiation result is based on ten independently established connections. The maximum relative sample standard deviations were 0.48\% for the connected baseline's steady-state components, 1.25\% for those of the proposed system, and 1.48\% for initiation overhead.

Steady-state active charge is the sum of the active contributions and excludes sleep. The 24-h and battery-lifetime models include each system's local sleep charge once per device and update cycle. This charge is calculated from the corresponding sleep current and actual update-cycle duration.

\subsubsection{Measured Current Profiles: Standard CS within an LE ACL Connection}
Bluetooth Core Specification v6.0 defines the Connection Interval (CI) from 7.5~ms to 4~s in 1.25~ms steps~\cite{b_core60}. While a longer CI reduces the charge consumed by connection events per update cycle, it increases latency for result transmission. RAS defines two Ranging Data exchange modes: Real-time Ranging Data (RT) and On-demand Ranging Data (OD). RT support is optional, whereas OD support is mandatory~\cite{b_ras}. In RT mode, the RAS Server sends results as they become available. In OD mode, it stores them for retrieval through the RAS Control Point.

To obtain a conservative comparison, all controllable choices relevant to the evaluation favored the connected baseline. It was derived from Nordic Semiconductor's nRF Connect SDK v3.1.0 RAS Initiator and Reflector samples~\cite{b_ncs_ras_initiator,b_ncs_ras_reflector}. In these samples, the Reflector hosts the RAS Server and reports Ranging Data to the Initiator, which processes them to obtain distance estimates. The Reflector was therefore selected for the primary comparison between the two systems as a constrained endpoint reporting to a fixed processing node, such as an anchor. The LE 2M PHY, an ATT MTU of 498~bytes, and a Link Layer data length of 251~bytes minimize airtime and fragmentation. Tx Power Dynamic Control enabled the transmit power to be set explicitly for each LE ACL connection. All transport, timing, and power settings were verified at runtime. Logging and debug output were disabled on the measured Reflector. The parameter values and low-power settings are listed in the Appendix.

To combine low initiation latency with low recurring connection event charge, the baseline uses separate CIs for initiation and steady-state operation. Connection establishment and the subsequent security, transport negotiation, RAS setup, and CS configuration use $\mathrm{CI}_{\mathrm{init}}=7.5~\mathrm{ms}$, the minimum supported by this implementation. Once setup is complete, the connection switches to a mode-specific steady-state CI $\mathrm{CI}_{\mathrm{ss}}$. This update is confirmed before CS is enabled.

In our tests, delivering all Ranging Data between consecutive CS procedures required two connection events in RT mode and seven in OD mode for both channel maps. The one-event RT and six-event OD alternatives did not reliably complete the transfer. For the 1-s update-cycle target, the longest compatible $\mathrm{CI}_{\mathrm{ss}}$ values are therefore 500~ms for RT and 142.5~ms for OD, giving actual update cycles of 1~s and 997.5~ms, respectively.

\begin{figure}[!t]
\centering
\includegraphics[width=\columnwidth]{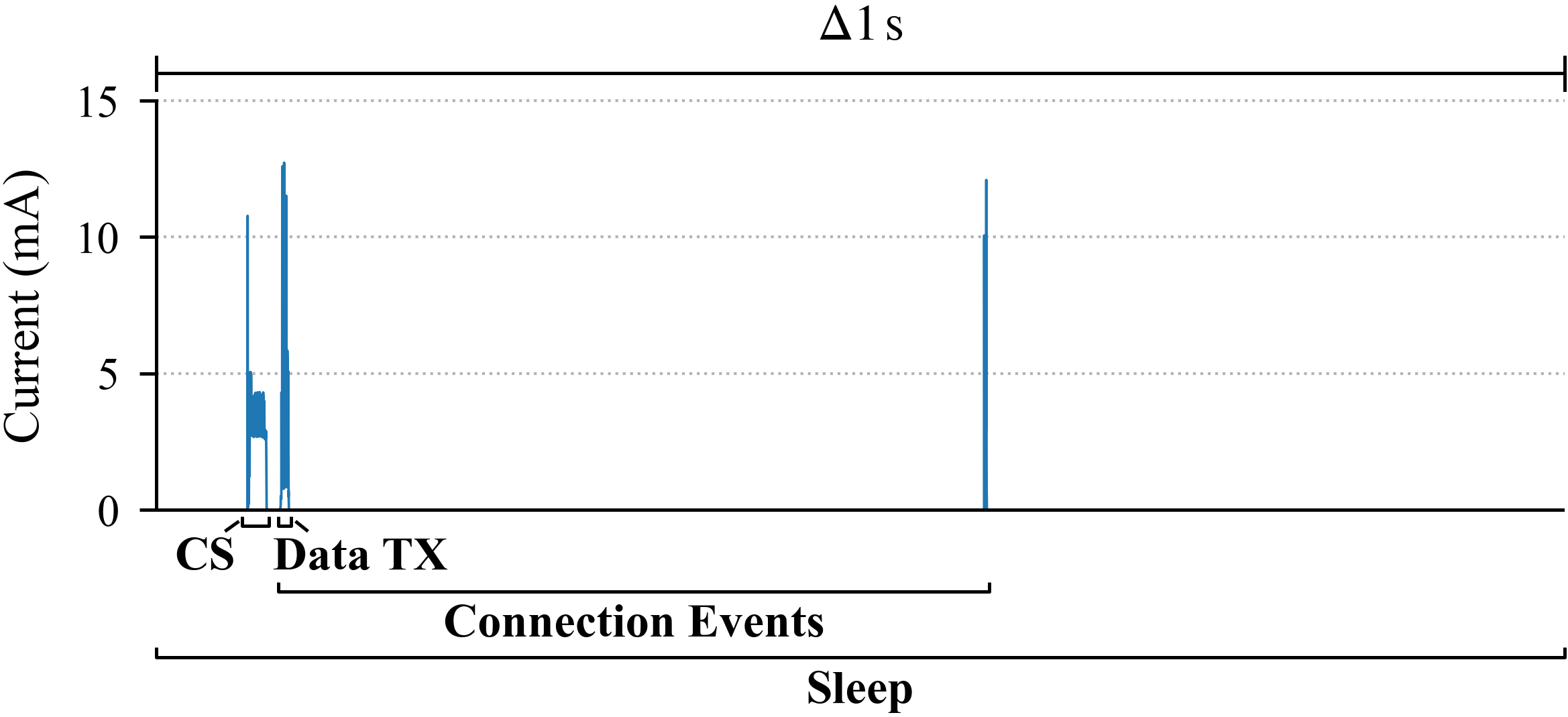}
\caption{Measured Reflector current profile of standard CS within an LE ACL connection in RT mode (1$\times$72~ch., $\mathrm{CI}_{\mathrm{ss}}=500~\mathrm{ms}$, $T_{\mathrm{upd}}=1~\mathrm{s}$). The CS burst is followed by a \emph{Data TX} connection event. The next connection event carries no Ranging Data.}
\label{fig:current_std_rt}
\end{figure}

\begin{figure}[!t]
\centering
\includegraphics[width=\columnwidth]{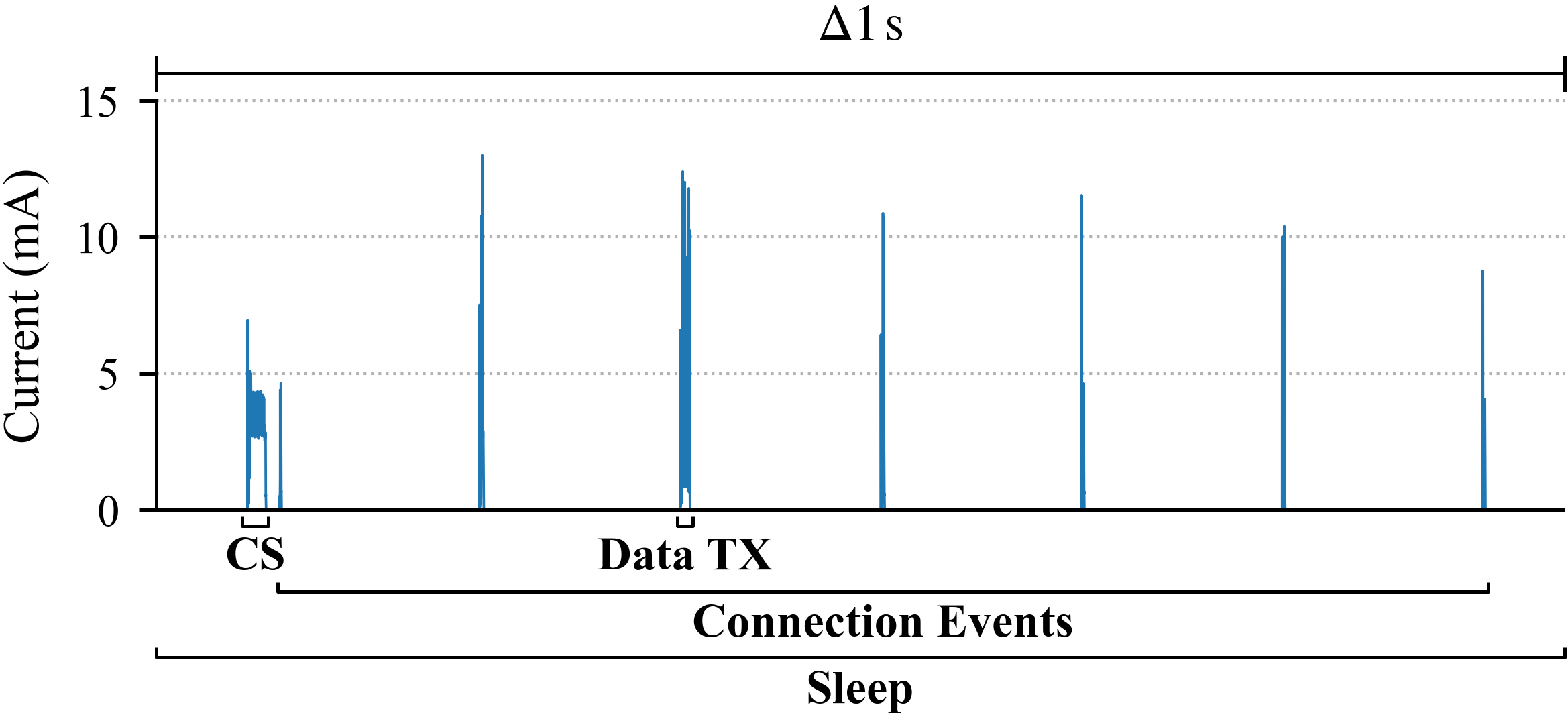}
\caption{Measured Reflector current profile of standard CS within an LE ACL connection in OD mode (1$\times$72~ch., $\mathrm{CI}_{\mathrm{ss}}=142.5~\mathrm{ms}$, $T_{\mathrm{upd}}=997.5~\mathrm{ms}$). Following the CS burst, retrieval and acknowledgment through the RAS Control Point and the Ranging Data transfer span multiple connection events. Ranging Data are transferred in the event marked \emph{Data TX}.}
\label{fig:current_std_od}
\end{figure}

Compared with RT mode, OD mode requires additional connection events for retrieval and acknowledgment through the RAS Control Point (see Fig.~\ref{fig:current_std_rt} and Fig.~\ref{fig:current_std_od}). This increases the charge per update cycle reported in Table~\ref{tab:energy_std}. The charge reported under \emph{Conn.\ events} accounts only for connection events that carry no CS measurement data. \emph{Data TX} represents the charge of connection events in which Ranging Data segments are transferred.

\begin{table}[!t]
\caption{Mean integrated charge of the Reflector in the connected baseline per update cycle with one partner ($K=1$). RT: $\mathrm{CI}_{\mathrm{ss}}=500~\mathrm{ms}$, $T_{\mathrm{upd}}=1~\mathrm{s}$. OD: $\mathrm{CI}_{\mathrm{ss}}=142.5~\mathrm{ms}$, $T_{\mathrm{upd}}=997.5~\mathrm{ms}$}
\label{tab:energy_std}
\centering
\begin{tabular*}{\columnwidth}{@{\hskip\tabcolsep\extracolsep{\fill}}lrrrrr@{\hskip\tabcolsep}}
\toprule
\multirow{2}{*}{\shortstack[l]{\textbf{RAS}\\\textbf{mode}}} & \multirow{2}{*}[-0.6ex]{\textbf{Channels}} & \multicolumn{4}{c}{\textbf{Charge ($\boldsymbol{\mu}$C)}} \\
\cmidrule(lr){3-6}
  & & \textbf{CS} & \textbf{Data TX} & \textbf{Conn.\ events} & \textbf{Sleep} \\
\midrule
RT & 37 & 25.18 & 12.12 & 3.27  & 3.05 \\
RT & 72 & 42.45 & 22.72 & 3.27  & 3.08 \\
OD & 37 & 23.98 & 14.61 & 23.78 & 2.25 \\
OD & 72 & 40.69 & 26.85 & 23.79 & 2.24 \\
\bottomrule
\end{tabular*}
\end{table}

\subsubsection{Measured Current Profiles: Proposed Connectionless CS via PAwR}
\begin{figure}[!t]
\centering
\includegraphics[width=\columnwidth]{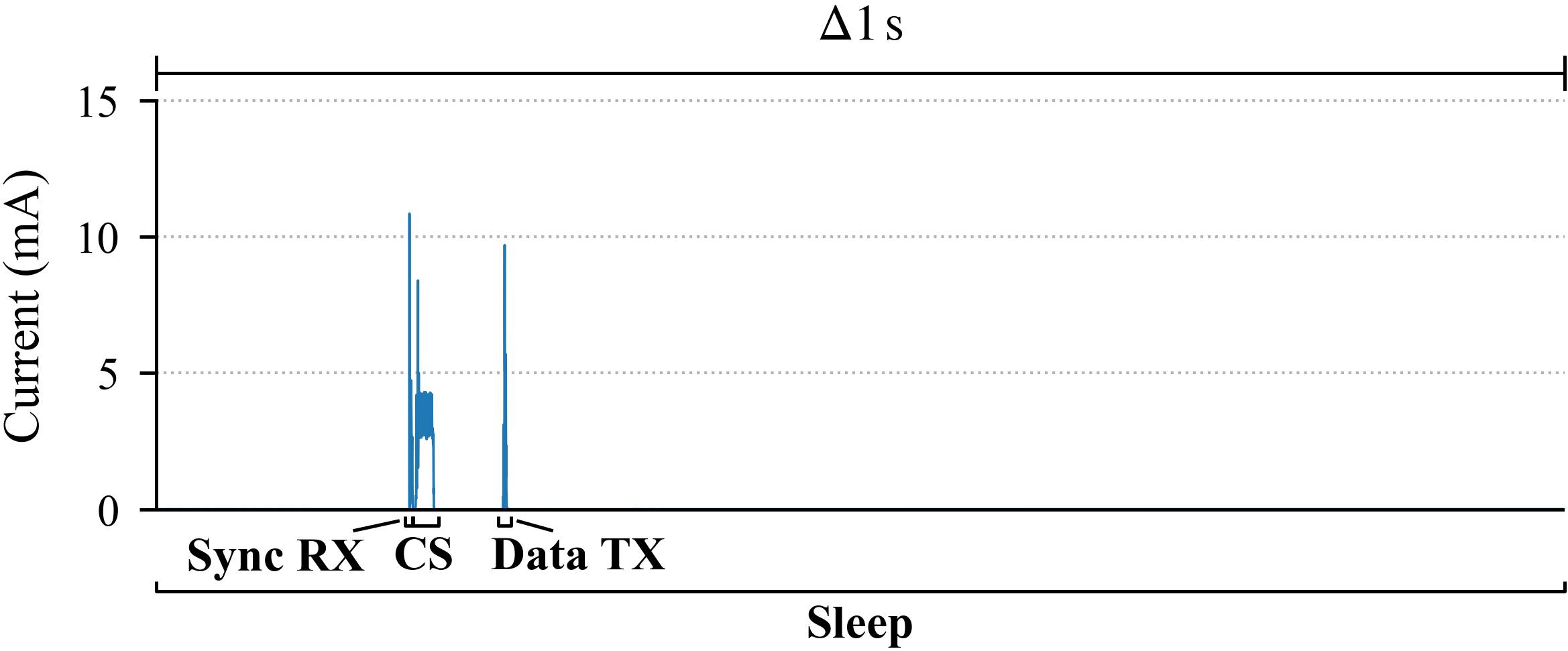}
\caption{Measured Reflector current profile of the proposed connectionless CS via PAwR (1$\times$72~ch., $T_{\mathrm{upd}}=1~\mathrm{s}$). The cycle comprises a brief synchronization reception, a CS burst, and a single response transmission.}
\label{fig:current_prop}
\end{figure}

The measured current profile in Fig.~\ref{fig:current_prop} shows that the proposed connectionless approach requires no periodic connection maintenance. For the single-partner energy measurements, the response slot delay was configured to 65~ms for both channel configurations. The synchronized CS device receives an \texttt{AUX\_SYNC\_SUBEVENT\_IND} PDU every update cycle. Without a measurement configuration update, the PDU carries no \texttt{AdvData}, and the corresponding synchronization charge is $Q_{\mathrm{sync}}=2.57~\mu\mathrm{C}$ (\emph{Sync RX} in Table~\ref{tab:energy_prop}). When an update is distributed, the PDU instead carries the new measurement configuration as a serialized configuration payload in its \texttt{AdvData} field. Receiving and processing this payload increases the charge to $Q_{\mathrm{cfg}}=6.9~\mu\mathrm{C}$. This value was measured with a fully occupied 249-byte \texttt{AdvData} field in a single PDU, providing a conservative basis for the partner-switching comparison.

\begin{table}[!t]
\caption{Mean integrated charge of the proposed connectionless CS via PAwR, measured at the Reflector per update cycle with one partner ($K=1$, $T_{\mathrm{upd}}=1~\mathrm{s}$)}
\label{tab:energy_prop}
\centering
\begin{tabular*}{\columnwidth}{@{\hskip\tabcolsep\extracolsep{\fill}}rrrrr@{\hskip\tabcolsep}}
\toprule
\multirow{2}{*}[-0.6ex]{\textbf{Channels}} & \multicolumn{4}{c}{\textbf{Charge ($\boldsymbol{\mu}$C)}} \\
\cmidrule(lr){2-5}
& \textbf{Sync RX} & \textbf{CS} & \textbf{Data TX} & \textbf{Sleep} \\
\midrule
37 & 2.57 & 25.40 & 4.23 & 2.57 \\
72 & 2.57 & 42.03 & 6.69 & 2.51 \\
\bottomrule
\end{tabular*}
\end{table}

\subsubsection{Steady-State Comparison}
To isolate the architectural differences between the two approaches, we compare their steady-state active charge per update cycle.

For the connected baseline, each partner requires its own LE ACL connection, giving
\begin{equation}
Q_{\mathrm{std,ss}}(K,\mathrm{CI}_{\mathrm{ss}}) = K\bigl[Q_{\mathrm{cs,std}} + Q_{\mathrm{data,std}} + Q_{\mathrm{conn}}(\mathrm{CI}_{\mathrm{ss}})\bigr],
\label{eq:qstd_ss}
\end{equation}
where $K$ is the number of distinct partners measured once per update cycle, $\mathrm{CI}_{\mathrm{ss}}$ is the steady-state CI in milliseconds, and $Q_{\mathrm{cs,std}}$, $Q_{\mathrm{data,std}}$, and $Q_{\mathrm{conn}}(\mathrm{CI}_{\mathrm{ss}})$ are the per-partner \emph{CS}, \emph{Data TX}, and \emph{Conn.\ events} charges in Table~\ref{tab:energy_std}.

For the proposed connectionless approach,
\begin{equation}
Q_{\mathrm{prop,ss}}(K) = Q_{\mathrm{sync}} + K\bigl[Q_{\mathrm{cs,prop}} + Q_{\mathrm{data,prop}}\bigr],
\label{eq:qprop_ss}
\end{equation}
where $Q_{\mathrm{sync}}$ is the per-cycle \emph{Sync RX} charge, and $Q_{\mathrm{cs,prop}}$ and $Q_{\mathrm{data,prop}}$ are the per-partner \emph{CS} and \emph{Data TX} charges in Table~\ref{tab:energy_prop}. Here, \emph{Data TX} denotes a PAwR response slot transmission.

The lower active charge in RT mode makes it the primary reference for the comparison in Table~\ref{tab:steady_state}. The savings stem mainly from two factors. First, the compact payload serialization described in Section~\ref{sec:data_plane} reduces \emph{Data TX} charge by 65\%--71\% relative to RT. Second, the proposed approach eliminates periodic connection events, whose larger charge contribution in OD mode leads to greater overall savings.

With multiple partners, the proposed approach also amortizes a single PAwR synchronization over all measurements. For four fixed partners (e.g., four anchors in a conventional localization deployment), the reduction reaches 28\% relative to RT at 72 channels and 51\% relative to OD at 37 channels.

\begin{table}[!t]
\caption{Reduction in steady-state active charge per update cycle ($T_{\mathrm{upd}}\approx1~\mathrm{s}$) relative to the connected baseline, computed from~\eqref{eq:qstd_ss} and~\eqref{eq:qprop_ss}. $K$ is the number of distinct partners measured once per update cycle}
\label{tab:steady_state}
\centering
\begin{tabular*}{\columnwidth}{@{\hskip\tabcolsep\extracolsep{\fill}}rrrr@{\hskip\tabcolsep}}
\toprule
\multirow{2}{*}[-0.6ex]{$\boldsymbol{K}$} & \multirow{2}{*}[-0.6ex]{\textbf{Channels}} & \multicolumn{2}{c}{\textbf{Reduction (\%)}} \\
\cmidrule(lr){3-4}
& & \textbf{RT} & \textbf{OD} \\
\midrule
1 & 37 & 21 & 48 \\
1 & 72 & 25 & 44 \\
4 & 37 & 25 & 51 \\
4 & 72 & 28 & 46 \\
\bottomrule
\end{tabular*}
\end{table}

Supplemental measurements of the proposed Initiator with one partner yielded 3\%--5\% lower steady-state active charge than the proposed Reflector across the two channel spacing configurations. The primary comparison therefore uses the role with the higher active charge in the proposed system.

\subsubsection{Initiation Overhead and Partner Switching}
Partner switching adds charge beyond steady-state operation. In connection-based operation, replacing a partner involves establishing a new LE ACL connection, repeating the initiation procedure, and tearing down the old connection. With frequent partner switching, initiation can dominate the connected baseline's total charge. By contrast, the proposed architecture treats partner switching as a measurement configuration update distributed via PAwR.

\begin{figure}[!t]
\centering
\includegraphics[width=\columnwidth]{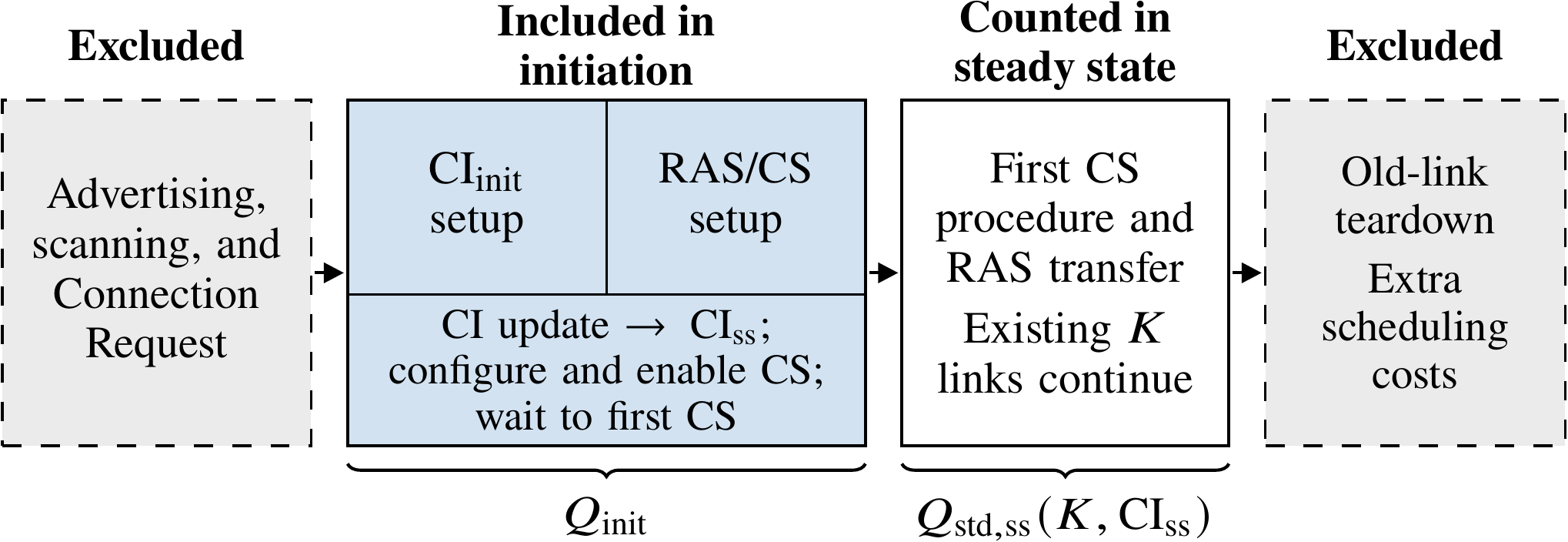}
\caption{Connected partner-switching accounting boundary. $Q_{\mathrm{init}}$ integrates the current above the local sleep baseline from the first activity of the established LE ACL connection to immediately before the first CS radio burst. The following CS procedure and RAS transfer contribute to $Q_{\mathrm{std,ss}}(K,\mathrm{CI}_{\mathrm{ss}})$. Phases marked \emph{Excluded} are omitted from the charge model.}
\label{fig:switching_timeline}
\end{figure}

To favor the connected baseline, we assume make-before-break operation, preparing each new link while existing measurement links remain active. This requires an additional free connection context and sufficient scheduling look-ahead. We count only the measured initiation charge $Q_{\mathrm{init}}$ and omit the additional connection-management and scheduling costs shown in Fig.~\ref{fig:switching_timeline}.

For a single partner switch, the proposed system requires only $Q_{\mathrm{cfg}}=6.9~\mu\mathrm{C}$ to receive and process the new measurement configuration. Compared with the connected baseline's measured initiation charges in Table~\ref{tab:init_overhead}, this is approximately 97\% less in either RAS mode.

\begin{table}[!t]
\caption{Mean initiation charge of the Reflector in the connected baseline (37 channels, $\mathrm{CI}_{\mathrm{init}}=7.5~\mathrm{ms}$)}
\label{tab:init_overhead}
\centering
\begin{tabular*}{\columnwidth}{@{\hskip\tabcolsep\extracolsep{\fill}}lrr@{\hskip\tabcolsep}}
\toprule
\textbf{RAS mode} & $\boldsymbol{\mathrm{CI}_{\mathrm{ss}}}$ \textbf{(ms)} & $\boldsymbol{Q_{\mathrm{init}}}$ \textbf{($\boldsymbol{\mu}$C)} \\
\midrule
RT & 500 & 224.31 \\
OD & 142.5 & 237.24 \\
\bottomrule
\end{tabular*}
\end{table}

To extend this comparison over a time horizon $T_{\mathrm{hor}}$, we model the charge for a device measuring $K$ partners per update cycle. Since synchronization reception is already included in $Q_{\mathrm{prop,ss}}$, one measurement configuration update per switching instant adds only $\Delta Q_{\mathrm{reconf}}=Q_{\mathrm{cfg}}-Q_{\mathrm{sync}}\approx4.33~\mu\mathrm{C}$ beyond the proposed system's steady-state charge. The total consumed charge is then
\begin{align}
Q_{\mathrm{std,tot}} &= N_{\mathrm{cyc}}\bigl[Q_{\mathrm{std,ss}}(K,\mathrm{CI}_{\mathrm{ss}}) + Q_{\mathrm{sleep,std}}\bigr] \nonumber \\
&\quad + N_{\mathrm{sw}}N_{\mathrm{rep}}Q_{\mathrm{init}}, \label{eq:qstd_total} \\[6pt]
Q_{\mathrm{prop,tot}} &= N_{\mathrm{cyc}}\bigl[Q_{\mathrm{prop,ss}}(K) + Q_{\mathrm{sleep,prop}}\bigr] \nonumber \\
&\quad + N_{\mathrm{sw}}\Delta Q_{\mathrm{reconf}}, \label{eq:qprop_total}
\end{align}
where $N_{\mathrm{cyc}}=T_{\mathrm{hor}}/T_{\mathrm{upd}}$ is the number of update cycles, $N_{\mathrm{sw}}$ is the number of switching instants, $N_{\mathrm{rep}}$ is the number of partners replaced at each instant, and $Q_{\mathrm{sleep,std}}$ and $Q_{\mathrm{sleep,prop}}$ are the respective local sleep charges per device and update cycle.

To show how switching frequency affects total charge, Table~\ref{tab:switch_example} applies the model to three switching scenarios over 24~h.

\begin{table}[!t]
\caption{Modeled total charge including sleep over 24~h for four partners ($K=4$), 37 channels, and $T_{\mathrm{upd}}=1~\mathrm{s}$, computed via~\eqref{eq:qstd_total} and~\eqref{eq:qprop_total}. The connected baseline uses RT with $\mathrm{CI}_{\mathrm{ss}}=500~\mathrm{ms}$ and $\mathrm{CI}_{\mathrm{init}}=7.5~\mathrm{ms}$. Each of the $N_{\mathrm{sw}}$ switching instants replaces all four partners ($N_{\mathrm{rep}}=4$)}
\label{tab:switch_example}
\centering
\begin{tabular*}{\columnwidth}{@{\hskip\tabcolsep\extracolsep{\fill}}lrrrr@{\hskip\tabcolsep}}
\toprule
\multirow{2}{*}[-0.6ex]{\textbf{Scenario}}
  & \multirow{2}{*}[-0.6ex]{$\boldsymbol{N_{\mathrm{sw}}}$}
  & \multicolumn{2}{c}{\textbf{Charge (C)}}
  & \multirow{2}{*}[-0.6ex]{\shortstack{\textbf{Reduction}\\\textbf{(\%)}}} \\
\cmidrule(lr){3-4}
  & & \shortstack{\textbf{Connected}\\\textbf{baseline}} & \textbf{Proposed} & \\
\midrule
No switching & 0 & 14.29 & 10.68 & 25 \\
Every 10~s & 8640 & 22.04 & 10.72 & 51 \\
Every update & 86\,400 & 91.81 & 11.06 & 88 \\
\bottomrule
\end{tabular*}
\end{table}

As switching becomes more frequent, initiation overhead accounts for a larger share of the connected baseline's total charge. Replacing all four partners every 10~s gives a 51\% charge reduction, rising to 88\% when they are replaced at every update. Including the connected costs excluded in Fig.~\ref{fig:switching_timeline} would further increase these savings, which are therefore lower bounds.

\subsubsection{Battery Lifetime Estimation}
To translate these charge savings into a practical lifetime estimate, we consider the proposed system powered by a CR2032 coin cell with a nominal capacity of 240~mAh at 3~V. The average current over an update cycle is
\begin{equation}
\bar{I} = \frac{Q_{\mathrm{active}}}{T_{\mathrm{upd}}} + I_{\mathrm{sleep}},
\label{eq:iavg}
\end{equation}
where $\bar{I}$ is the average current in microamperes, $Q_{\mathrm{active}}$ is the active charge per update cycle in microcoulombs, $T_{\mathrm{upd}}$ is the update-cycle duration in seconds, and $I_{\mathrm{sleep}}$ is the locally measured sleep current of approximately $2.57~\mu\mathrm{A}$. For longer $T_{\mathrm{upd}}$, the active charge is amortized over a longer interval while the sleep-current term remains constant. The corresponding lifetime is
\begin{equation}
L = \frac{1000\,C}{24\,\bar{I}},
\label{eq:lifetime}
\end{equation}
where $L$ is the lifetime in days and $C$ is the nominal battery capacity in milliampere-hours.

For the four-partner scenario with 37 channels, we assume one measurement configuration update per cycle. The measured components give $Q_{\mathrm{active}}=Q_{\mathrm{cfg}}+4\,(Q_{\mathrm{cs,prop}}+Q_{\mathrm{data,prop}})\approx125.42~\mu\mathrm{C}$. With $T_{\mathrm{upd}}=1~\mathrm{s}$, the average current is approximately $128~\mu\mathrm{A}$, giving an estimated lifetime of approximately 3~months. At $T_{\mathrm{upd}}=30~\mathrm{s}$, the average current decreases to approximately $6.75~\mu\mathrm{A}$, extending the estimated lifetime to over 4~years.

\subsection{System Scalability and Capacity}
\label{sec:scalability}

\subsubsection{Timing Model}
\begin{figure}[!t]
\centering
\includegraphics[width=\columnwidth]{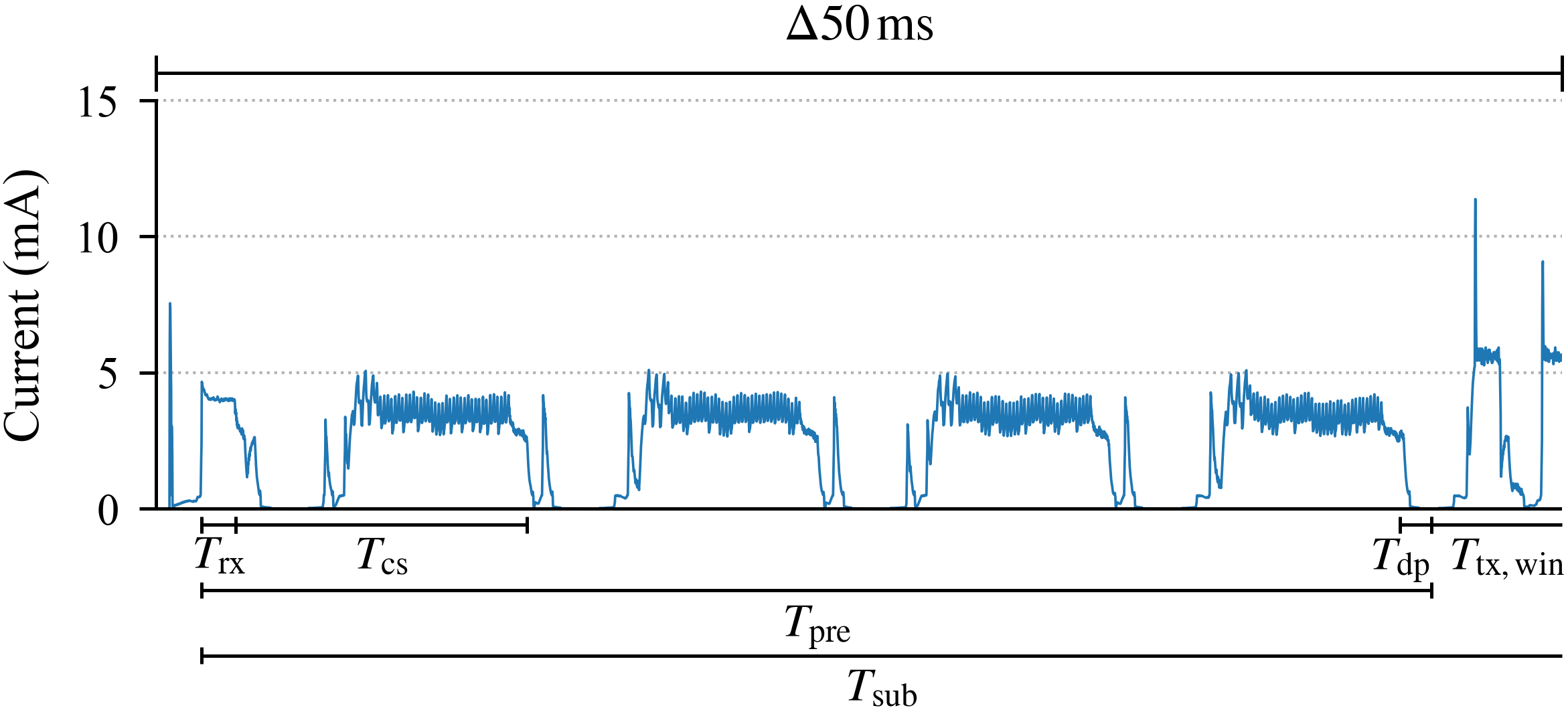}
\caption{Measured subevent timing phases for the representative four-partner workload ($K=4$) over 37 channels. The complete wake-up peak before the \texttt{\bfseries AUX\_SYNC\_SUBEVENT\_IND} PDU lies outside the marked $T_{\mathrm{rx}}$, which begins with PDU reception. Four contiguous $T_{\mathrm{cs}}$ intervals represent the effective durations of the subsequent CS procedures, followed by data processing during $T_{\mathrm{dp}}$. The pretransmission phase $T_{\mathrm{pre}}$ comprises these intervals and equals the response slot delay. The two visible bursts within the response slot window $T_{\mathrm{tx,win}}$ correspond to $N_{\mathrm{rsp}}=2$ response PDUs. Open right brackets indicate that $T_{\mathrm{tx,win}}$ and the full subevent $T_{\mathrm{sub}}$ extend beyond the displayed interval.}
\label{fig:timing_model}
\end{figure}

Within one PAwR subevent, an active device receives the PAwR subevent PDU, executes its configured CS procedures, processes the resulting data, and transmits one or more response PDUs in the response slot window. Fig.~\ref{fig:timing_model} shows the corresponding measured current profile annotated with these timing phases.

The pretransmission duration combines reception, the configured CS procedures, and data processing,
\begin{equation}
T_{\mathrm{pre}} = T_{\mathrm{rx}} + N_{\mathrm{meas}}\,T_{\mathrm{cs}} + T_{\mathrm{dp}},
\label{eq:t_pre}
\end{equation}
\begingroup
\emergencystretch=1em
where $T_{\mathrm{rx}}$ is the duration from the start of PDU reception to the first $T_{\mathrm{cs}}$ interval, $T_{\mathrm{cs}}$ is the effective duration associated with one CS procedure, $T_{\mathrm{dp}}$ is the data-processing delay between the final $T_{\mathrm{cs}}$ interval and the response slot window, and $N_{\mathrm{meas}}$ is the number of CS procedures per device per update cycle. In PAwR terms, $T_{\mathrm{pre}}$ corresponds to the response slot delay, i.e., the configured offset (an integer multiple of 1.25~ms) between the start of the \texttt{AUX\_SYNC\_SUBEVENT\_IND} PDU and the first response slot~\cite{b_core60}. The response slot window contains $N_{\mathrm{rs}}$ slots of spacing $T_{\mathrm{rs}}$,
\begin{equation}
T_{\mathrm{tx,win}} = N_{\mathrm{rs}}\,T_{\mathrm{rs}}.
\label{eq:t_txwin}
\end{equation}
The total subevent duration is then
\begin{equation}
T_{\mathrm{sub}} = T_{\mathrm{pre}} + T_{\mathrm{tx,win}}.
\label{eq:t_sub}
\end{equation}
\par
\endgroup

The Bluetooth Core Specification limits the subevent interval, and with it $T_{\mathrm{sub}}$, to at most $T_{\mathrm{sub}}^{\max}=318.75~\mathrm{ms}$ and the number of response slots per subevent to at most 255~\cite{b_core60}. Because the response slot delay and the response slot window must fit within the subevent interval, the configurable slot count is bounded by
\begin{equation}
N_{\mathrm{rs}} \;\leq\; \min\!\left(255,\; \left\lfloor \frac{T_{\mathrm{sub}}^{\max} - T_{\mathrm{pre}}}{T_{\mathrm{rs}}} \right\rfloor\right).
\label{eq:n_rs_max}
\end{equation}
For the representative four-partner workload at $T_{\mathrm{rs}}=1.25~\mathrm{ms}$, the subevent interval is the binding constraint and admits up to 220 response slots for the 37-channel and 204 for the 72-channel configuration.

Table~\ref{tab:timing_model} separates the measured pretransmission terms from the modeled durations of the response slot window and full subevent. $T_{\mathrm{cs}}$ increases with the number of channels because additional mode-2 steps extend each CS procedure, whereas $T_{\mathrm{rx}}$ and $T_{\mathrm{dp}}$ remain nearly constant. $T_{\mathrm{tx,win}}$ and $T_{\mathrm{sub}}$ follow from the block-aligned response slot counts derived below.

\begin{table}[!t]
\caption{Measured pretransmission terms and modeled durations of the response slot window and full subevent for the representative four-partner workload ($K=4$, $N_{\mathrm{AP}}=1$, $T_{\mathrm{rs}}=1.25~\mathrm{ms}$). Modeled durations use the largest block-aligned response slot count permitted by~\eqref{eq:n_rs_max}}
\label{tab:timing_model}
\centering
\begin{tabular*}{\columnwidth}{@{\hskip\tabcolsep\extracolsep{\fill}}lrrrrrr@{\hskip\tabcolsep}}
\toprule
\multirow{2}{*}[-0.6ex]{\textbf{Channels}} & \multicolumn{4}{c}{\textbf{Measured (ms)}} & \multicolumn{2}{c}{\textbf{Modeled (ms)}} \\
\cmidrule(lr){2-5}\cmidrule(lr){6-7}
 & $\boldsymbol{T_{\mathrm{rx}}}$ & $\boldsymbol{T_{\mathrm{cs}}}$ & $\boldsymbol{T_{\mathrm{dp}}}$ & $\boldsymbol{T_{\mathrm{pre}}}$ & $\boldsymbol{T_{\mathrm{tx,win}}}$ & $\boldsymbol{T_{\mathrm{sub}}}$ \\
\midrule
37 & 1.21 & 10.35 & 1.14 & 43.75 & 275 & 318.75 \\
72 & 1.21 & 15.29 & 1.39 & 63.75 & 250 & 313.75 \\
\bottomrule
\end{tabular*}
\end{table}

\subsubsection{Capacity and Update-Rate Bounds}
When every device is active in all $N_{\mathrm{ms}}$ measurement slots, each device requires $N_{\mathrm{rsp}}$ response slots per update cycle, as defined in~\eqref{eq:n_rsp}. The paired response slot allocation uses the minimum block size of $2N_{\mathrm{rsp}}$ slots for each two-device allocation group. The maximum subevent group size is therefore
\begin{equation}
N_{\mathrm{dev}} \;\leq\; N_{\mathrm{dev}}^{\max} = 2\left\lfloor \frac{N_{\mathrm{rs}}}{2N_{\mathrm{rsp}}} \right\rfloor.
\label{eq:devs_per_sub}
\end{equation}
With up to $N_{\mathrm{sub}}=128$ subevents per PAwR train, the maximum total number of active devices across all subevent groups is
\begin{equation}
N_{\mathrm{dev,tot}}^{\max} = N_{\mathrm{sub}}\,N_{\mathrm{dev}}^{\max}.
\label{eq:cap_total}
\end{equation}
Because every subevent group is served sequentially within the train, the per-device update cycle is bounded below by the train period:
\begin{equation}
T_{\mathrm{upd}} \;\geq\; T_{\mathrm{upd}}^{\min} = N_{\mathrm{sub}}\,T_{\mathrm{sub}}.
\label{eq:tupd_min}
\end{equation}
Collision-free parallel operation within a subevent group uses the deterministic channel management introduced in Section~\ref{sec:coordination}. In the capacity model, the cyclic-shift construction is applied independently to complementary even and odd subsets containing 37 and 35 of the 72~physical channels, respectively, raising the channel-side limit to 72 simultaneous pairs. The measured 37-channel timing provides a conservative bound for both subsets. Capacity is therefore determined by the response slot budget through~\eqref{eq:devs_per_sub}, with $N_{\mathrm{rs}}$ limited by~\eqref{eq:n_rs_max}.

For the representative four-partner workload, all four measurement slots are active and address distinct partners, so $K=N_{\mathrm{meas}}=N_{\mathrm{ms}}=4$. At 37 channels and 2~MHz spacing, two per-path results are concatenated per response payload ($k=2$, see Section~\ref{sec:data_plane}). Together with the prototype's single antenna path ($N_{\mathrm{AP}}=1$, see Table~\ref{tab:config_params}), this gives $N_{\mathrm{rsp}}=2$. The paired allocation therefore uses four-slot blocks, making all $N_{\mathrm{rs}}=220$ slots available. Each subevent group can contain $N_{\mathrm{dev}}^{\max}=110$ active devices, yielding a total modeled capacity of 14\,080~devices and a minimum update cycle of 40.8~s. At 72 channels and 1~MHz spacing, $k=1$ gives $N_{\mathrm{rsp}}=4$. The resulting eight-slot blocks and longer $T_{\mathrm{pre}}$ leave $N_{\mathrm{rs}}=200$ slots, yielding $N_{\mathrm{dev}}^{\max}=50$ active devices per subevent group and 6400~devices in total at a minimum update cycle of 40.16~s.

Because both modeled subevent durations are at or near their maximum, the longer pretransmission phase at 72 channels reduces the response slot budget while both minimum update cycles remain close to 40~s. Practical deployments can choose a point on the capacity-latency tradeoff by tuning $N_{\mathrm{ms}}$, $N_{\mathrm{meas}}$, channel count, $N_{\mathrm{rs}}$, and $N_{\mathrm{sub}}$ to the application-specific update-rate budget.

These bounds use the shortest CS step timings supported by the SoftDevice Controller in the evaluated nRF Connect SDK v3.1.0 implementation~\cite{b_ncs_sdc_cs}, as listed in Table~\ref{tab:config_params}. The Bluetooth Core Specification permits shorter values for some of these parameters~\cite{b_core60}. Such timings would directly reduce $T_{\mathrm{cs}}$ and hence $T_{\mathrm{pre}}$, freeing response slots for additional capacity within the same subevent interval.

\section{Conclusion and Future Work}
\label{sec:conclusion}
This paper presented a connectionless BLE CS architecture that combines the LE CS Test command with PAwR to orchestrate scalable and energy-efficient ranging without per-pair LE ACL connections. Centrally distributed measurement configurations support arbitrary device-to-device pairings and allow synchronized CS devices to derive role, DRBG state, channel sequence, and response slot assignment locally. Compact serialization enables complete result reporting through PAwR response slots. The proof-of-concept evaluation shows that deterministic channel management eliminates the collision-induced outliers observed under simulated dense-deployment channel overlaps. For one measured partner per 1-s update cycle, the proposed approach reduces steady-state active charge by up to 25\% relative to the fair connected baseline in RT mode and by up to 48\% in OD mode. The charge advantage increases under partner switching because receiving and processing a new configuration payload requires approximately 97\% less charge than the connected initiation procedure. When all four partners are replaced in each update cycle, the component-based 24-h model consequently projects up to 88\% lower total charge. The architecture thus enables energy-efficient ranging beyond the conventional tag-to-anchor pattern, with arbitrary, dynamically reconfigurable device-to-device pairings. For the representative four-measurement workload with a single antenna path, the timing-based capacity analysis projects a collision-free capacity of up to 14\,080 active devices per PAwR train at a minimum update cycle of approximately 40~s. The application's required update rate therefore constrains capacity.

Future work at the system level targets multi-GW synchronization and dynamic device assignment to subevent groups. Redundant, synchronized GWs could eliminate the GW as a single point of failure and enable CS procedures between devices served by different GWs. Dynamic assignment would complement the supported within-group partner switching by relaxing the fixed group boundaries that currently prevent cross-group measurements. Initial assignments could use coarse spatial cues, whereas accumulated pairwise distance estimates could guide later regrouping by spatial proximity. Channel management could also combine a shared pseudorandom permutation with adaptive channel exclusion following the principle of Adaptive Frequency Hopping. A common CS channel map distributed by the CO would preserve same-step collision avoidance within the remaining channel set while reducing exposure to persistent and time-varying interference.

At the processing level, advanced estimators could improve CS ranging in multipath-rich indoor environments. Fusion with inertial measurements could improve the robustness of downstream localization. Aggregating the pairwise distance estimates into a cooperative measurement graph could extend the architecture from ranging to indoor localization, supporting both anchor-free relative localization and anchor-supported absolute positioning across the deployment.

At the protocol level, an optional application-layer security extension with preprovisioned keys could authenticate measurement configurations and support a keyed derivation of \texttt{DRBG\_Nonce}, mitigating its predictability under LE CS Test with $K_{\mathrm{DRBG}}=0$. Since the development of this prototype, Bluetooth Core Specification v6.3 has introduced CS Inline PCT Transfer (IPT)~\cite{b_core63}. An IPT-enabled mode could omit Reflector response payloads, reducing response slot demand, the charge of PAwR response transmissions, and aggregation latency. Deterministically rotating the Initiator role, e.g., based on the event counter, would distribute the remaining reporting load across CS devices.

\appendix[Configuration Parameters]
\section*{Appendix\\*Configuration Parameters}
\label{sec:appendix_config}

Table~\ref{tab:config_params} lists the CS, Bluetooth Controller, and system parameters used in the evaluation, followed by settings specific to the connected baseline.

\begin{table}[!t]
\caption{Configuration parameters of the experimental evaluation using nRF Connect SDK v3.1.0}
\label{tab:config_params}
\centering
\begin{tabular*}{\columnwidth}{@{\hskip\tabcolsep\extracolsep{\fill}}p{0.38\columnwidth}p{0.46\columnwidth}@{\hskip\tabcolsep}}
\toprule
\textbf{Parameter} & \textbf{Value} \\
\midrule
\multicolumn{2}{l}{\textit{CS Parameters}} \\
Main Mode & Mode 2 \\
Mode-0 Steps & 3 \\
Mode-2 Channels & {\raggedright 72 (1~MHz spacing)\newline 37 (2~MHz spacing)\par} \\
Antenna Paths ($N_{\mathrm{AP}}$) & 1 \\
CS Synchronization PHY & LE 2M \\
CS Procedure PHY & LE 2M \\
Channel Map Repetition & 1 \\
CS TX Power & {\raggedright $+8$~dBm (Section~\ref{sec:collision})\newline 0~dBm (Sections~\ref{sec:energy} and~\ref{sec:scalability})\par} \\
$T_{\mathrm{IP1}}$ & 60~$\mu$s \\
$T_{\mathrm{IP2}}$ & {\raggedright 30~$\mu$s (Section~\ref{sec:collision})\newline 20~$\mu$s (Sections~\ref{sec:energy} and~\ref{sec:scalability})\par} \\
$T_{\mathrm{FCS}}$ & 60~$\mu$s \\
$T_{\mathrm{PM}}$ & 10~$\mu$s \\
$T_{\mathrm{SW}}$ & 0~$\mu$s \\
\midrule
\multicolumn{2}{l}{\textit{Bluetooth Controller Parameters}} \\
Data PHY & LE 2M \\
Data TX Power & 0~dBm \\
\midrule
\multicolumn{2}{l}{\textit{System Parameters}} \\
System Timer & GRTC \\
Tickless Kernel & enabled \\
Low-Frequency Clock & LFXO \\
Low-Power State & System ON \\
Retained RAM & 256~KB \\
\midrule
\multicolumn{2}{l}{\textit{Connected Baseline Parameters}} \\
RAS Modes & RT and OD \\
Initial CI ($\mathrm{CI}_{\mathrm{init}}$) & 7.5~ms \\
Steady-State CI ($\mathrm{CI}_{\mathrm{ss}}$) & RT: 500~ms; OD: 142.5~ms \\
LE Security Level & 2 \\
Bonding & disabled \\
Unenhanced ATT Bearers & 1 \\
Enhanced ATT & disabled \\
$\mathrm{ATT\_MTU}$ & 498~bytes \\
Tx/Rx Data Length & 251~bytes \\
Connection Event Extension & enabled \\
Tx Power Dynamic Control & enabled \\
Ranging Data Buffers & 2 per LE ACL connection \\
\bottomrule
\end{tabular*}
\end{table}

\section*{Data and Code Availability}
The measurement data and analysis scripts supporting the experimental results reported in this article are available from the corresponding author upon reasonable request.

\bibliographystyle{IEEEtran}
\bibliography{refs}

\begingroup
\makeatletter
\def\@IEEEBIOskipN{2\baselineskip}
\makeatother
\begin{IEEEbiography}[{\includegraphics[width=1in,height=1.25in,clip,keepaspectratio]{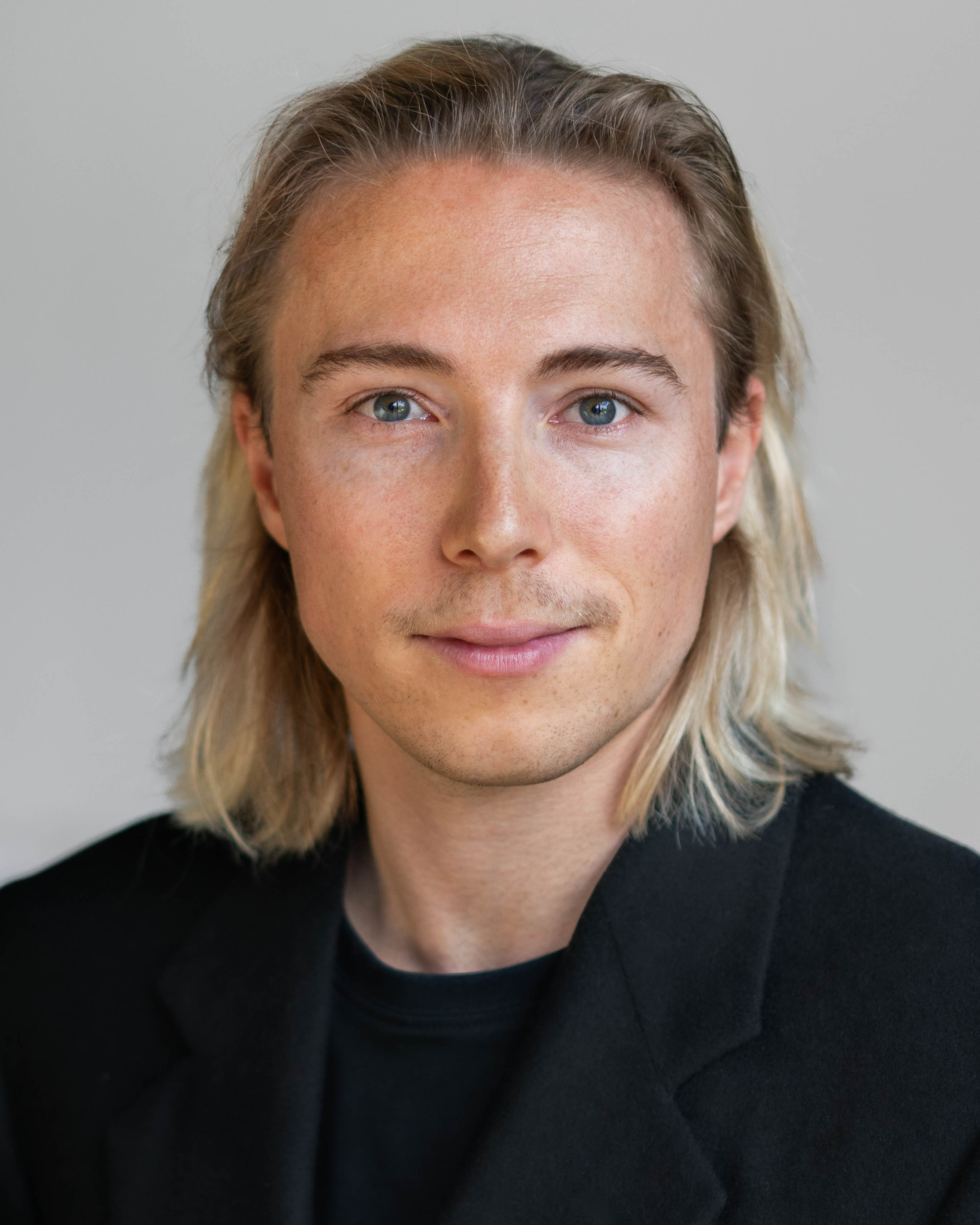}}]{Leon Schex}
(Graduate Student Member, IEEE) received the M.Sc.\ degree in computer science and systems engineering from TH Köln -- University of Applied Sciences, Cologne, Germany, and the M.Sc.\ degree in intelligent systems and applications from Université Gustave Eiffel, Marne-la-Vallée, France, both in 2025 within a double-degree program.

He is currently pursuing the Ph.D. degree with TH Köln -- University of Applied Sciences, Cologne, Germany, where he is a Research Associate with the Embedded Real-Time Systems Lab (ERTS·LAB), Institute of Computer and Communication Technology. His research interests include Bluetooth Channel Sounding, indoor localization and ranging, scalable and energy-efficient wireless sensor networks, signal processing, machine learning, and sensor fusion.
\end{IEEEbiography}

\begin{IEEEbiography}[{\includegraphics[width=1in,height=1.25in,clip,keepaspectratio]{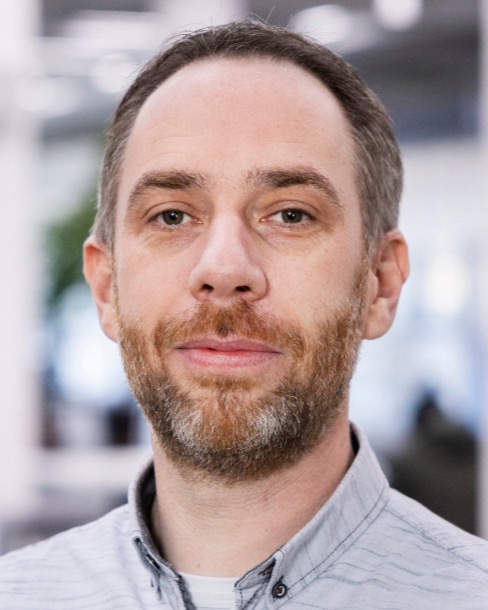}}]{Markus Cremer}
received the Ph.D. degree in telecommunications engineering from London South Bank University, London, U.K., in 2016.

He is currently a Professor of Embedded Real-Time Systems with the Faculty of Information, Media and Electrical Engineering, TH Köln -- University of Applied Sciences, Cologne, Germany, where he leads the Embedded Real-Time Systems Lab (ERTS·LAB) and serves as program director for the B.Sc.\ program in Electrical Engineering and Information Technology. His research interests include indoor localization and real-time locating systems (RTLS), currently focusing on Bluetooth Channel Sounding and ultra-wideband (UWB) positioning, as well as sensor fusion, with applications spanning industrial systems and animal behavior monitoring.
\end{IEEEbiography}

\vfill\newpage
\begin{IEEEbiography}[{\includegraphics[width=1in,height=1.25in,clip,keepaspectratio]{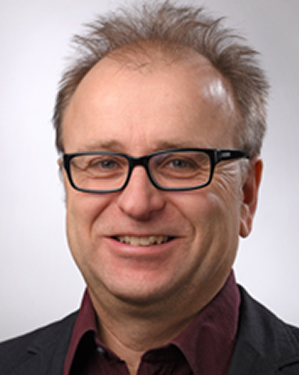}}]{Uwe Dettmar}
(Member, IEEE) received the Dipl.-Ing.\ degree in electrical engineering in 1989 and the Dr.-Ing.\ degree in 1994, both from the Technical University of Darmstadt, Darmstadt, Germany.

He subsequently worked for four years in industrial research laboratories at Ascom, Switzerland, and Ericsson, Germany, focusing on wireless technology and radio access. Since 1999, he has been a full Professor at TH Köln -- University of Applied Sciences, Cologne, Germany, where he leads the Digital Communications and IoT Laboratory within the Institute of Computer and Communication Technology. Throughout his tenure, he has led numerous publicly and privately funded research projects. His research interests include communication system design, localization and positioning, sensor networks for the Internet of Things (IoT), and error-correcting codes.

Dr.\ Dettmar is author and coauthor of over 50 patents, conference papers, and journal articles. He is a member of the VDE.
\end{IEEEbiography}

\EOD
\endgroup
\end{document}